\documentclass[12pt]{article}
\pdfoutput=1
\usepackage{epsfig}
\usepackage{graphicx}
\usepackage{cite}
\usepackage{amsfonts}
\usepackage{amssymb}
\usepackage{bm}
\usepackage{latexsym}
\usepackage{amsmath}
\numberwithin{equation}{section}
\def\ee{\end{equation}}
\def\ba{\begin{eqnarray}}
\def\ea{\end{eqnarray}}

\def\bq{\begin{quote}}
\def\eq{\end{quote}}

 at 10truept
\newcommand{\bfx}{{\bf x}}

\newcommand{\bfk}{{\bf k}}

\newcommand{\beq}{\begin{equation}}
\newcommand{\eeq}{\end{equation}}
\newcommand{\beqa}{\begin{eqnarray}}
\newcommand{\eeqa}{\end{eqnarray}}
\newcommand{\bea}{\begin{eqnarray}}
\newcommand{\eea}{\end{eqnarray}}
\newcommand{\p}{\partial}

 \newcommand{\ep}{\epsilon}

\newcommand{\vect}[1]{\bm{\mathrm{{#1}}}}

\newcommand{\khat}{{\hat q}}

\newcommand{\Hotps}{\left(\frac{H}{2\pi}\right)^2}

\def\lesssim{~\mbox{\raisebox{-.6ex}{$\stackrel{<}{\sim}$}}~}

\def\ltap{\ \raise.3ex\hbox{$<$\kern-.75em\lower1ex\hbox{$\sim$}}\ }
\def\gtap{\ \raise.3ex\hbox{$>$\kern-.75em\lower1ex\hbox{$\sim$}}\ }
\def\gl{\ \raise.5ex\hbox{$>$}\kern-.8em\lower.5ex\hbox{$<$}\ }
\def\roughly#1{\raise.3ex\hbox{$#1$\kern-.75em\lower1ex\hbox{$\sim$}}}
\def\calo{{\cal O}}
\def\kq{{\kappa_q}}
\def\kqi{{\kappa_{q,i}}}

\def\kk{{\kappa_k}}
\def\cals{{\cal S}}
\def\adot{{\dot a}}
\def\phidot{{\dot \phi}}

\begin{document}

\thispagestyle{empty}
\begin{titlepage}
\nopagebreak

\title{
Fluctuating geometries, q-observables, and infrared growth in inflationary spacetimes} 

\vfill
\author{Steven B. Giddings$^{a}$\footnote{giddings@physics.ucsb.edu}\ \  and Martin S. Sloth$^{b,c}$\footnote{sloth@cern.ch}
}
\date{ }


\maketitle

\vskip 0.5cm

{\it  $^{a}$ Department of Physics, 
University of California, Santa Barbara, CA 93106}

\vskip 0.5cm
{\it  $^{b}$
D\'{e}partment de Physique Th\'{e}orique and Center for Astroparticle Physics,\\
$^{}~~~~~$ Universit\'{e} de Gen\`{e}ve, 24 Quai E. Ansermet, CH-1211 Gen\`{e}ve, Switzerland}

\vskip 0.5cm
{\it  $^{c}$ CERN, Physics Department, Theory Unit, CH-1211 Geneva 23, Switzerland}

\vfill
\begin{abstract}
Infrared growth of geometrical fluctuations in inflationary spacetimes is investigated.  The problem of gauge-invariant characterization of growth of perturbations, which is of interest also in other spacetimes such as black holes, is addressed by studying evolution of the lengths of curves in the geometry.  These may either connect freely falling ``satellites," or wrap non-trivial cycles of geometries like the torus, and are also used in diffeomorphism-invariant constructions of two-point functions of field operators.  For spacelike separations significantly exceeding the Hubble scale, no spacetime geodesic connects two events, but one may find geodesics constrained to lie within constant-time spatial slices.  In inflationary geometries, metric perturbations produce significant and growing corrections to the lengths of such geodesics, as we show in both quantization on an inflating torus and in standard slow-roll inflation.   These become large, signaling breakdown of a perturbative description of the geometry via such observables, and consistent with perturbative instability of de Sitter space.  In particular, we show that the geodesic distance on constant time slices during inflation becomes non-perturbative a few e-folds after a given scale has left the horizon, by distances $\sim 1/H^3\sim R S$, obstructing use of such geodesics in constructing IR-safe observables based on the spatial geometry.  We briefly discuss other possible measures of such geometrical fluctuations.
 \end{abstract}
 \vskip.4in
 
\noindent
CERN-TH-PH/2011-187\hfill \\  
\vfill
\end{titlepage}

\setcounter{equation}{0} \setcounter{footnote}{0}

\section{Introduction}

A long-standing problem has been formulation of a more complete quantum treatment of inflationary cosmology.  Gravity has familiar short-distance problems ({\it e.g.} renormalizability, singularities), but also {\it long-distance} problems that become evident in inflationary cosmology, as well as in black holes.  Those include apparent infrared (IR) growth of perturbations (see {\it e.g.} \cite{Linde:2005ht}), the general problem of specifying gauge-invariant observables (see {\it e.g.} \cite{Giddings:2005id}), and, in the former case, related problems of measures\footnote{For a review see \cite{Freivogel:2011eg}.}.

Understanding the IR growth of perturbations, particularly of the ``gas" of gravitons produced in inflation, has in part been confounded by the challenges of formulating observables sensitive to these perturbations.  Interpretations of this growth has ranged from the claim that through backreaction they cause decay of the cosmologocal constant \cite{Polyakov:1982ug,Polyakov:2009nq,Polyakov:2007mm,Tsamis:1992sx,Tsamis:1993ub,Tsamis:1994ca,Tsamis:2007is,Myhrvold:1983hu}, to the belief that they have little meaningful effect.  What is needed is a gauge-invariant way to describe physical effects of such fluctuations.

Here it is important to note a distinction:  if we imagine that there is a yet-unknown complete quantum-mechanical description of cosmology, then one of its basic features should be a set of quantum-mechanical observables. However, it is clear that, due to our limitations as Earth-bound observers in a particular era of cosmology, there is a much more restricted set of quantities that we can actually measure -- and that an observational cosmologist would describe as ``observable."  Much of the discussion of this paper focusses on the former notion of observable.  Where it is confusing, the former may be distinguished by calling them q-observables\cite{Giddings:2011zd}: a given q-observable may or may not be observable to Earth's cosmologists. 

One goal of this paper is to formulate candidate q-observables sensitive to fluctuations in geometry.  The latter in particular can have variance that becomes large at large times, and so a related goal is to better understand the meaning of this growth.  The considerations of this paper are more formal than those of \cite{Giddings:2011zd}, which discussed possible observable (to us) effects.  These are related since an observable must ultimately be a q-observable.  Moreover, there is common problem with both (related also to the measure problem) and that is to provide quantities that are ``infrared safe," and thus well-defined even in the presence of large IR fluctuations.  In \cite{Giddings:2011zd}, infrared-safe observables were defined, relevant for  the observations of a late-time observer of the CMB sky, by gauge transforming away metric fluctuations with wavelengths outside the horizon of the observer; a similar prescription was investigated in  \cite{HHH,HMM}.\footnote{In particular, in \cite{HMM} it was similarly argued that one could apply such a gauge transformation, removing long wavelength modes. Their specific transformation leads to an exponential truncation of long wavelength modes given essentially by multiplying the IR part of the spectrum by a factor $1-e^{-\rho k}$ with $\rho>0$. For arbitrary choice of $\rho$ this does not correspond to the spectrum of the gauge invariant comoving curvature perturbation, in terms of which late-time observables are usually expressed by cosmologists \cite{wood}. However if $\rho$ is identified with the size of the observed volume, then the proposal is essentially similar to the proposal of \cite{Giddings:2011zd}, with the only difference that \cite{Giddings:2011zd} applied a step function instead of a smooth exponential window function. In the language of \cite{Giddings:2011zd}, $\rho$ corresponds to a renormalization scale (in \cite{Giddings:2011zd} labelled $1/q$), set by the choice of observer.}

An alternate approach, in particular pursued in \cite{Byrnes:2010yc,Urakawa:2010it,Urakawa:2010kr,Urakawa:2011fg,Gerstenlauer:2011ti}, is 
that  correlation functions  written in terms of the unperturbed geometric distance on the reheating surface would constitute candidate IR-safe observables.  This is {\it not} necessarily what a late-time observer measures, but is an interesting prescription to consider.  
Part of the problem with such a construction is that the curvature perturbation is not conserved when written in terms of the geodesic distance of the reheating surface. But even worse, as we show in section \ref{SDCM}, when fluctuations of the geodesic distance are taken into account, this distance receives large perturbative corrections, and the description of the geometry in terms of geodesic distance becomes problematic. This happens only a few e-folds after a given scale has exited the horizon.

In contrast, in terms of the comoving momentum the curvature perturbation is conserved on large scales, which makes the correlation functions of the curvature perturbation in comoving momentum at horizon crossing a convenient variable for discussing the observables for a late-time observer after the end of inflation. For this correlation function the effect of longer wavelength fluctuations, exiting the horizon even earlier, is to shift the background of the shorter wavelength mode when it exits the horizon. This can be described in terms of the semi-classical relations developed in \cite{Giddings:2010nc}. To a late-time observer, who can measure the correlation function in many different patches of the sky, this becomes a physical effect. For a very late-time observer with access to a very large inflated volume set by of order $N\sim 1/H^2$ e-folds (or a duration of order $t\sim R_{dS}S_{dS}$, where $R_{dS}$ and $S_{dS}$ are the de Sitter radius and entropy respectively) of inflation, this effect becomes non-perturbative, while a present day observer with only access to the last $60$ e-folds of inflation will see only a small effect on the CMB sky. A renormalization group equation relates the different observers with different survey volumes\cite{Giddings:2011zd}.

Despite the drawbacks from the viewpoint of present-day observation, q-observables based on spacetime geometry have some promise for giving gauge-invariant descriptions of inflation.  For that reason, this paper will describe construction of q-observables of this kind, and investigate what they tell us about the physics of inflation.  A specific focus will be on q-observables capable of probing growing variance, {\it e.g.} of tensor fluctuations, and interpreting its meaning for the geometry of inflation.  In short, while we cannot support arguments for decay of the cosmological constant, rapid expansion produces buildup of long distance modes leading to large cumulative fluctuations after sufficiently long time.  So, the growing variance {\it does} appear to indicate a quantum instability of inflationary spacetimes such as de Sitter (dS) space, to large excursions from the semiclassical geometry, and indicates a breakdown of our perturbative treatment.  Indeed, this behavior is apparently another facet of the basic phenomenon responsible for self-reproducing inflation \cite{Giddings:2010nc}.

In outline, the next section describes in more detail the problem of observables in inflation, and the question of defining q-observables using geometrical measures such as lengths of geodesics.  In particular, we note that the latter cannot be formulated based on spacetime geodesics spanning super-horizon distances, though one may base constructions on geometry of specific time slicings.  We turn to study of these in subsequent sections.  Section 3 sets up perturbative quantization of slow-roll inflation, in different gauge-fixings.  Section 4 describes a basic construction, where one diagnoses properties of a fluctuating geometry by the lengths of curves in it.  We note that growing metric fluctuations make important contributions, but also that care must be exercised in order to find gauge-invariant quantities.  In particular, an arbitrary curve is not gauge invariant, and as a consequence it is natural to consider instead curves that are {\it geodesics} of a fluctuating geometry.

A simple context to examine such curves, and effects of metric fluctuations on them, is on a spatial torus $T^3$, where we use a cycle of the torus to avoid the problem of specifying endpoints.  Quantization on a toroidal version of dS is described.  We find that tensor fluctuations do have a significant effect on torus cycles, giving a gauge-invariant statement of large corrections at large times, which we take to be a diagnostic of large fluctuations in the geometry of the spacetime.  Section 6 then considers geodesics with endpoints, which may be thought of as quantum particles or classical ``satellites," or as locations of field operators, in gauge-invariant constructions.  We find similar large contributions of metric fluctuations, due to fluctuations of the geodesic in the geometry.  These effects become particularly pronounced in slow-roll.  This section also describes various subleading effects, quantum and otherwise, in such a construction.  The large contributions of metric fluctuations to geodesic lengths signal breakdown of our perturbative description of the geometry via such q-observables.   Section 7 considers the related problem of fluctuations in geometrical volume, making contact with \cite{Creminelli:2008es,Dubovsky:2008rf}.  
Section 8 ends with brief description of other possible q-observable diagnostics of quantum inflationary geometry, and with conclusions.  An appendix contains additonal details on gauge fixing.

\section{Q-observables and tensor fluctuations}\label{tensfluc}

\subsection{Challenges of gauge invariance}

We seek gauge invariant observables in inflationary universes, particularly characterizing geometry.  Such gauge-invariant observables are surprisingly difficult to formulate in quantum gravity, where the symmetry of the effective field theory description is diffeomorphism invariance.  This means that local observables familiar from field theory are not gauge invariant, and that observables must take a more non-local form.  One would in particular like to formulate such observables that reduce to the local observables of field theory in an approximation; general aspects of this problem, and proposals for such ``proto-local" observables, are described in \cite{Giddings:2005id}.

\begin{figure}[!hbtp] \label{fig1}
\begin{center}
\includegraphics[width=5cm, angle=270]{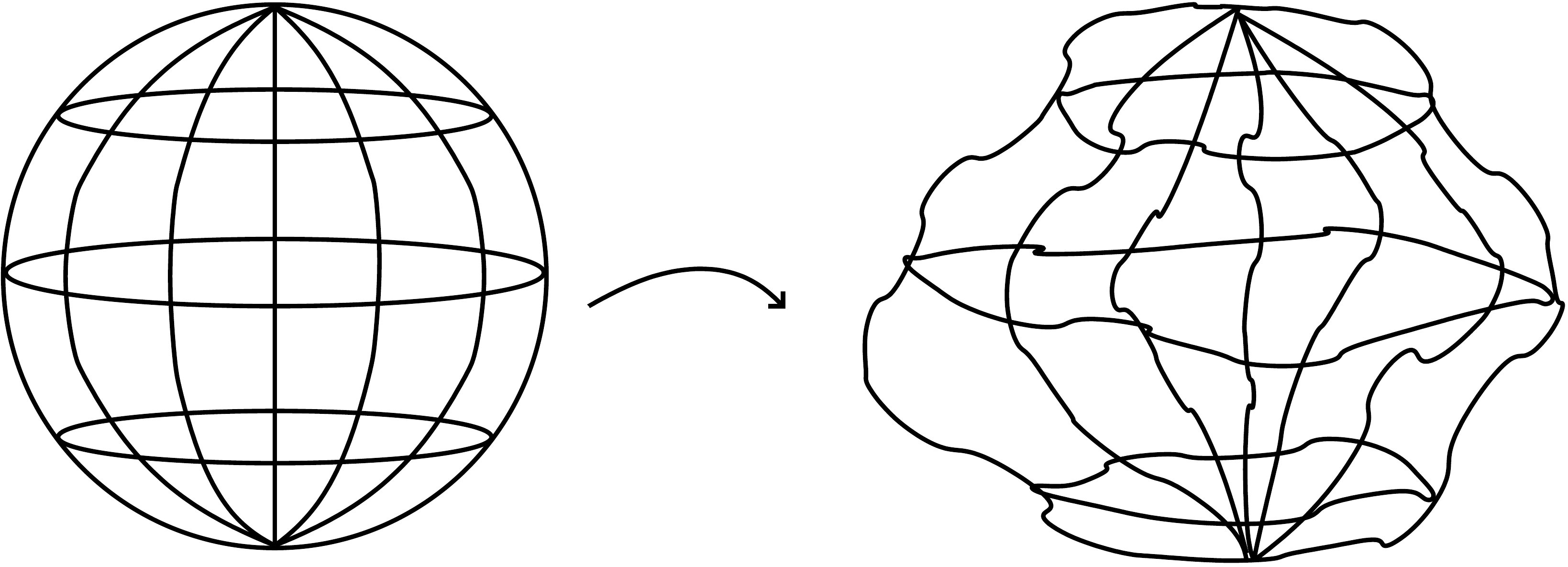}
\end{center}
\caption{Metric perturbations can lead to significantly different geometries, as illustrated for the sphere.}
\end{figure}

A focus of this paper will be on observables sensitive to fluctuations of geometry, and specifically to their apparent growth as evidenced {\it e.g.} by growth in variance of the tensor modes of the graviton.  Here there are various puzzles. In particular, the growing variance is associated with tensor fluctuations that exit the horizon scale and are redshifted to long wavelength.  However, at least locally a long-wavelength metric fluctuation is unobservable.  Consider, for example, a small perturbation in a cosmological metric, with scale factor $a(t)$,
\beq\label{pertmet}
ds^2=-dt^2 +a^2(t)(e^{\Gamma})_{ij}dx^idx^j\ ,
\eeq 
where $\Gamma_{ij}$ is a spacetime-dependent metric perturbation (in notation of  \cite{Giddings:2011zd}), which generally contains both scalar and traceless tensor parts.

In the long-wavelength limit, $\Gamma_{ij}$ is constant, and can be removed by a change of coordinates:
\beq\label{constxm}
x^i \rightarrow (e^{-\Gamma/2})_{ij} {\tilde x}^j\ .
\eeq 
Indeed, this is reflected if one attempts to formulate observables in terms of local scalar quantities, such as the curvature; the curvature due to the perturbation vanishes in the limit of large redshift.  This fact was also used in \cite{Giddings:2011zd}, where it was argued that IR-safe quantities relevant to describing observations in a given observer's horizon can be formulated by using a transformation such as \eqref{constxm} to eliminate the effect of longer-wavelength fluctuations.

However, there is a different viewpoint indicating that there is physics in long-wavelength perturbations of the gravitational field.  In particular, the long-wavelength regime (compared to the Planck scale) is precisely the regime that gravitational wave detectors such as LIGO and Virgo are designed to probe.  These detectors span kilometer sizes precisely because they are not measuring local quantities, but instead integrated quantities, specifically the relative displacement of two test masses separated by a finite distance.  This physical picture suggests that more ``global" observables are important in gravity.  

Indeed, consider spatial slices in an inflating spacetime; focus concretely on cases of closed slices, such as $S^3$ or $T^3$.  While it is true that the perturbations redshift as the slices expand, in order to compare two different slices one needs to perform a rescaling so that they have the same size.  Once such a rescaling has been performed, growth of perturbations such as in \eqref{pertmet} appear to lead to growing ``lumpiness" of the space, as illustrated in fig.~1.

There are various quantities one could try to use to characterize such growth of lumpiness. One possibility is to use an experimental apparatus to define an area, and measure rotation of vectors upon parallel transport around its perimeter; this kind of ``Wilson loop" integral is sensitive to the curvature over the area's span.  While the curvature may fall with the scale factor as $1/a^2$, the area of such a loop, if comoving, has opposite scaling, so the rotation of the vector remains fixed under scaling.  One problem, however, is that specification of such a loop is not diffeomorphism-invariant, without additional physical structure.  One could likewise describe other observable quantities (volume, {\it etc.}), though with similar problems of gauge invariance.

An even more primitive kind of observable is the kind measured in gravitational-wave astronomy, simply the distance between points defined by heavy point masses.  These can be thought of as giving gauge-invariant ``reference points" for measuring geometry.

\subsection{Q-observables via geodesic lengths}

We will pursue such an approach to defining observables sensitive to tensor fluctuations in cosmology.  For example, one could consider a cosmology with a gas of massive test particles (or ``satellites") present, and give one characterization of the gravitational fluctuations in terms of fluctuations of the distances between the test particles; these distances are given in terms of integrals of the metric over paths connecting the particles.   

A more fundamental characterization of this problem takes into account the quantum nature of the particles.  
Indeed, a related class of observables have been proposed, in order to reproduce the two-point function in field theory.  
Let $\sigma$ be a scalar field, and define the two-point function at a fixed {\it geodesic} separation $\cals=s(x,x')$:
\beq
\label{corrFL}
\langle\sigma(x)\sigma(x')\rangle_{\big |s(x,x')=\cals}\ .
\eeq
Such a construction has been investigated in \cite{Ambjorn:1996wc,Giddings:2005id,Urakawa:2010it,Urakawa:2010kr,Urakawa:2011fg,Byrnes:2010yc,Gerstenlauer:2011ti}, and used in the context of the simplicial approach to quantum gravity\cite{Hamber}.   This non-local construction can also be given for other local operators, such as the curvature scalar $R(x)$.  

However, one immediately encounters an important limitation to characterizations of the geometry by lengths of geodesics, and of correlation functions via a construction like \eqref{corrFL}.  This arises from the statement that arbitrary spacelike-separated points cannot be connected by a spacelike geodesic in inflationary geometries, and is most easily illustrated in de Sitter space.  While we are familiar with the statement that in inflationary geometry, objects quickly become causally disconnected in the sense that there is no timelike or null geodesic connecting them, this property at first seems rather surprising in comparison to more familiar spacetimes.  A quick argument is the following: suppose two points are connected by a spacelike geodesic.  Then, by a dS symmetry transformation, the points and the geodesic may be taken to lie on the ``equator" (minimal radius sphere) of the global slicing.  But, no two points on this slice are separated by a distance greater than $\pi/H$, so this serves as an upper limit on the shortest geodesic distance between two points.

This property can be investigated in the flat slicing \eqref{pertmet}. Consider vanishing perturbation $\Gamma$, and, without loss of generality, a geodesic with tangent orthogonal to the $x^2=y$ and $x^3=z$ directions.  The $\Gamma=0$ metric \eqref{pertmet} has Killing vector 
\beq\label{kvect}
k^{\mu} = (0,1,0,0)~,
\eeq
and so $k_\mu dX^\mu/ds=c$
is conserved along a geodesic.  Introducing the conformal time $\eta$ through
\beq\label{conftime}
dt = a(t) d\eta\ ,
\eeq
conservation immediately gives the equation
\beq\label{curveeq}
d\eta^2 = dx^2\left(1-{a^2\over c^2}\right)\ .
\eeq
For de Sitter space, 
\beq\label{dsscale}
a=e^{Ht}=-1/(H\eta)\ ,
\eeq
and \eqref{curveeq} can be rewritten
\beq\label{curveeq2}
dx = {\eta d\eta\over\sqrt{\eta^2-\eta_0^2}}\ ,
\eeq
with solution
\beq
(x-x_0)^2 = \eta^2 - \eta_0^2\ .
\eeq
This curve reaches maximum time   at $\eta=\eta_0<0$ and $x=x_0$ .  Rewriting it as
\beq
H^2a^2(x-x_0)^2= 1- {a^2\over a_0^2}\ ,
\eeq
we find that two such points in a time slice with  $\eta<\eta_0$ have a maximum ``physical" distance measured in the slice given by 
\beq\label{xmax}
a\Delta x_{\rm max}=2/H\ , 
\eeq
from the curve with $a_0\rightarrow\infty$.  The geodesic distance is easily computed, and found to be
\beq\label{smax}
\cals = {\pi \over H} - {2\over H}\tan^{-1}\left({1\over \sqrt{(\eta/\eta_0)^2-1}}\right)\ ,
\eeq
with maximum $\cals_{\rm max}={\pi/ H}$.  Thus, for fixed {\it comoving} distance $\Delta x$, points rapidly exceed the bounds \eqref{xmax}, \eqref{smax}, just outside the horizon, and there is no geodesic connecting them.  This emphasizes the extreme causal and geometrical separation between points that is intrinsic to inflation.

If there are no geodesics connecting generic points in an inflationary geometry, one must seek other ways to characterize the geometry.  One approach which we will follow in this paper is to find geodesics subject to an additional constraint, {\it e.g.} that they lie in a given spatial slice, in a particular time slicing.\footnote{Another alternative is to consider curves with, {\it e.g.}, fixed acceleration.  We thank D. Marolf for discussions on such an approach.}  This requires some additional structure, to play the role of a ``clock."  One candidate is an evolving scalar field, though without a potential an initial field velocity redshifts away.  This does leave, however, the case of, {\it e.g.}, slow-roll inflation.  One can seek other candidates in the geometry itself, {\it e.g.} using the local expansion in the volume element as a clock.  Or, one may consider additional physical structure.

In a construction involving satellites or such as \eqref{corrFL}, one faces the question of disentangling the quantum dynamics of the particles from that of the gravitational field.  For this reason, we will first focus on what appears to be an even simpler problem, temporarily dispensing with the particle ``anchors" for the world-line ends, before returning to  discuss these later in the paper.  Namely, if we consider a classical de Sitter cosmology, in the flat slicing, an equally good solution results from periodic identification under translations in $R^3$ to give the the torus $T^3$.  In such a geometry, the proper lengths of the cycles of the torus, measured via the integrated distance, appear to be useful characteristics  of the geometry.  Passing to the quantum theory, we can consider gravitational fluctuations about our inflating torus, and see what their effect is on these length variables.  Of course a proper treatment of this problem involves examination of possible gauge-dependencies of these length variables.  However, we anticipate that this problem might be easier here, given that the curves are anchored to the homotopy of the torus, which is a topological invariant.

Specifically, describe the torus via the identification $x^i\sim x^i+L$.  We will first consider as a candidate q-observable the length around the x-cycle of $T^3$, given by the line integral
\beq \label{Win}
\cals = \oint ds = \oint dx \sqrt{g_{xx} } ~.
\eeq
The zeroth-order in $\Gamma$ result is $\cals=a(t)L$.  However, fluctuations in $\Gamma_{ij}$ will produce fluctuations in this length, and in particular we might anticipate significant corrections resulting from the infrared growth in these fluctuations.  In order to explore this question, we need a careful treatment of their quantization.

\section{Action, perturbative expansion, quantization}
\label{sec2}
\subsection{Action and perturbative expansion}
We begin with dynamical preliminaries. 
For concreteness we will focus on characterizing the geometry of models of single-field, slow-roll inflation, or of its ``no-roll" de Sitter limit. The action is
\beq\label{actions}
S=\frac{1}{2}\int d^4x \sqrt{-g}\left[{{\cal R}\over 8\pi G} - (\nabla \phi)^2 - 2V(\phi)\right]~,
\eeq
where ${\cal R}$ is the spacetime Ricci scalar.  In the de Sitter limit, $V$ is just the cosmological constant, $V\rightarrow\Lambda=$const.  

A perturbative description of the coupled metric and matter fluctuations can be derived\cite{Maldacena:2002vr}
using the Arnowitt--Deser--Misner\cite{Arnowitt:1960es} (ADM) parameterization of the line element,
\begin{equation}\label{admmetric}
    ds^2= -N^2 dt^2 + h_{ij}(dx^i + N^idt)(dx^j + N^jdt)~,
\end{equation}
where $N$, $N^i$ are the lapse and
the shift functions.  

We begin with a solution to the equations of motion,
\beq\label{classsoln}
N=1\quad,\quad N^i=0\quad,\quad h_{ij}= a^2(t)\delta_{ij}\quad,\quad \phi=\phi_0(t)\ ,
\eeq
with instantaneous Hubble scale $H={{\dot a}/ a}$, and
satisfying the slow-roll conditions
\beq
 \epsilon = (V')^2/2V^2\ll 1\quad,\quad \eta = V''/V \ll 1\ . 
 \eeq
  An arbitrary perturbation about this may be parameterized as
\beq\label{spacemet}
h_{ij} = a^2(t)e^{2\zeta} (e^{\gamma})_{ij}\quad,\quad \phi=\phi_0+\varphi\ ,
\eeq
together with $\Delta N = N-1$ and $N^i$, where $\gamma_{ij}$ is a traceless matrix, $\gamma_{ii}=0$.   (Here we adopt the convention of summing over lower indices with the flat-space metric $\delta_{ij}$.)

One requires a gauge condition fixing the diffeomorphism symmetry; different choices will be discussed below.  The lapse and shift act as Lagrange multipliers, whose equations of motion are the constraint part of Einstein's equations.  After a gauge is fixed, one may solve these equations perturbatively to determine $\Delta N$ and $N^i$ in terms of $h_{ij}$ and $\phi$.  As a result, the dynamical degrees of freedom consist of the two polarization modes of the graviton, contained in $\gamma_{ij}$, and one scalar, described by a gauge-dependent combination of $\zeta$ and $\phi$.    
In the de Sitter limit, this scalar decouples from the metric at quadratic order; alternately, with no field $\phi$, the ``time translation" symmetry implies that the 
scalar curvature perturbation $\zeta$ can be gauged away.  

One treats the fluctuations in a perturbation expansion in $1/M_p^2 = 8 \pi G$; we will typically work in units where $M_p=1$.  In the slow-roll regime, $H^2 \simeq V/3M_p^2$.

\subsection{Gauge fixing}

The gauge symmetries are diffeomorphisms, $x^\mu = x^{\mu\prime} +\epsilon^\mu$, under which the linear variation in the metric is
\beq\label{gendiff}
\delta g_{\mu\nu} = \nabla^{(0)}_\mu \epsilon_\nu + \nabla^{(0)}_\nu\epsilon_\mu\ .
\eeq
This implies that the variations of $\gamma_{ij}+2\zeta\delta_{ij}$ and $\phi$ take the form
\bea\label{gaugexm}
\delta(\gamma_{ij}+ 2\zeta\delta_{ij}) &=& \partial_i \epsilon_j + \partial_j \epsilon_i + 2 H \epsilon^0 \delta_{ij} +\calo\left(\epsilon^2, \epsilon\gamma, \epsilon\varphi, \epsilon N^i\ , \epsilon \Delta N\right)\ \cr
\delta \varphi &=& \epsilon^0 {\dot\phi_0}+\calo\left(\epsilon^2,\epsilon\varphi\right)\ ,
\eea
where subleading terms are higher-order in the $1/M_p$ expansion.  (Explicit formulas for these and higher order gauge-fixing are described in the appendix.)

The traceless perturbation, $\gamma_{ij}$, is independent of choice of time slice at this order.  The spatial diffeomorphisms are typically fixed by choosing transverse gauge,\footnote{Note that perturbations satisfying these conditions can be thought of as ``gauge invariant" (see {\it e.g.} \cite{Mukhanov:1990me}).}
\beq
\partial_i\gamma_{ij}=0\ .
\eeq
Specifically, a spatial diffeomorphism gives the transformation (where we are careful to extract the traceless piece)
 \beq\label{gTtrans}
\partial_i\gamma^{\prime}_{ij} = \partial_i \gamma_{ij} + \partial_i^2 \epsilon_j + {1\over 3} \partial_j \partial_i\epsilon_i + \calo\left(\epsilon^2, \epsilon\gamma, \epsilon N^i\ , \epsilon \Delta N\right)\ .
\eeq
Thus, at linear order, we can satisfy
\beq\label{transverse}
\partial_i\gamma^{\prime}_{ij}=0
\eeq
by choosing
\beq\label{transfix}
\epsilon_i = - \frac{\delta_{ij} -{1\over 4} \partial_i\partial_j/\nabla^2}{\nabla^2} \partial_k \gamma_{kj}\ ,
\eeq
where $1/\nabla^2$ is the Green function for $\nabla^2=\partial_i\partial_i$.
One can use the expression (\ref{transfix}) to fix the gauge (\ref{transverse}) at higher order in $1/M_p$ as well, by working order-by-order and including the extra contributions in (\ref{gTtrans}) on the right hand side of (\ref{transfix}) (see appendix).

Two commonly-used gauges for fixing the time-slicing are {\it comoving gauge},
\beq\label{constphi}
\varphi =0
\eeq
and {\it traceless} or {\it spatially-flat} gauge,
\beq\label{zerozeta}
\zeta=0\quad .
\eeq
As we see from (\ref{gaugexm}), these choices can be made order-by-order in an expansion in $1/M_p$. 
The choice \eqref{constphi} is very physical, in that constant-$\phi$ slices provide the time-slicing.  The alternate choice \eqref{zerozeta} can also be characterized physically, as the choice where the evolution of the local volume element is given by the original unperturbed solution.

Once the gauge is fixed by (\ref{transverse}) and either \eqref{constphi} or \eqref{zerozeta}, 
the lapse $N$ and shift $N^i$ are determined by the constraint equations.  
For example, in pure de Sitter, with the gauge \eqref{zerozeta}, and at linear order, these are
\beq
2H\p_jN^{(1)}-\frac{1}{2}\p^ 2N_j^{(1)}+\frac{1}{2}\p_j\p_iN_i^{(1)} =0
\eeq
and
\beq
\Delta N^{(1)} =-\frac{1}{3H}\p_iN_i^{(1)}~,
\eeq
which, when combined, imply
\beq\label{constcomb}
(\p^2\delta_{ij}+\frac{1}{3}\p_i\p_j)N_i^{(1)} =0~.
\eeq
Thus, a linear-order solution is $N^{(1)}=N_i^{(1)}=0$, and this procedure can be generalized to slow roll and higher-order contributions.

Note also that the transverse-traceless choice (\ref{transverse}) and \eqref{zerozeta} may not completely fix the gauge.  In particular, we see that at linear level these conditions are 
unaffected by residual gauge transformations satisfying
\beq\label{residxm}
(\p^2\delta_{ij}+\frac{1}{3}\p_i\p_j)\epsilon_i =0
\eeq
with a corresponding $\epsilon^0$ chosen to preserve $\gamma=0$.  The residual transformations are of course compatible with the constraints, as can be seen from the linear transformations
\beq\label{lingauge}
N\to N+\dot\ep^0\ , \ N^i \to N^i+\dot\ep^i +h^{ij}\p_i\ep_0
\eeq 
and comparing (\ref{constcomb}) and (\ref{residxm}).   The dS isometries provide such residual transformations.

\subsection{Quantization}

Once a gauge has been fixed, the action \eqref{actions} can be expanded in the parameter $H/M_p$, 
\beq
S=S_0 + S_2 + S_I\ .
\eeq
Here, $S_0$ is the action of the classical solution \eqref{classsoln}, and the first-order term vanishes by the equations of motion.  The second order terms are given in, for example, \cite{Maldacena:2002vr}.
For the tensors, we have in transverse gauge
\beq\label{tensact}
S_{2,t} = {1\over 8} \int   dt d^3 x a^3 \left[ ({\dot \gamma_{ij}})^2 - a^{-2} (\partial_k\gamma_{ij})^2\right]\ .
\eeq
The scalar action depends on the gauge.  For example, in the gauge \eqref{constphi}, one finds
\beq\label{scalact}
S_{2,s}={1\over 2} \int dt d^3x\, a^3 {{\dot \phi_0}^2a^2\over {\dot a}^2} \left[  {\dot \zeta}^2 - a^{-2} (\partial_i \zeta)^2\right]\ .
\eeq

Higher-order interaction terms described by $S_I$ can likewise be worked out, expanding in $\gamma+2\zeta$ and in $\varphi$.  In doing so, one solves the constraints for $\Delta N$ and $N^i$, order-by-order in terms of the other degrees of freedom; these then contribute to the interaction terms.

Correlators are then computed perturbatively in these interactions, with the leading, gaussian, correlators determined by the actions \eqref{tensact}, \eqref{scalact}.  In particular, these correlators exhibit a growing variance at long times; for example the two-point function for tensors, averaged over the proper length scale $a/H$ to regulate the UV, behaves as \cite{Starobinsky:1979ty}
\beq\label{Tensvar}
\langle \gamma^2(x)\rangle = \frac{1}{4}\langle \gamma_{ij}(x) \gamma_{ij}(x)\rangle\approx 2\frac{H^2}{(2\pi)^2}\log\left[\frac{a(t)} {a_i}\right]\ ,
\eeq
where $a_i$ gives an IR cutoff corresponding to a ``beginning" of inflation.

\section{Fluctuating line integrals}

It is well known that the variance of fluctuations of light fields in de Sitter can become large, but, as explained in the introduction, an important question regards the gauge-invariant or observable consequences of this statement.  Naively, one might expect the inflated space to be very smooth.   But, further reflection suggests that the accumulation of fluctuations at zero physical momentum, driven by the expansion, could have an important effect, connected to the growing variance \eqref{Tensvar}.  In particular, pile up of these fluctuations suggests that the spacetime could have a very inhomogeneous structure at the longest scales.

We will study such inhomogeneities using the kinds of q-observable described in section \ref{tensfluc}, namely via integrated lengths of curves, which are basic characteristics of any geometry.  Given the issues with unconstrained (four-dimensional) geodesics described in section \ref{tensfluc}, our focus will be on geodesics within the three-dimensional geometries corresponding to constant-time slices in either of the gauges \eqref{constphi}, \eqref{zerozeta}.  Our basic picture is that space becomes lumpy on large scales (see fig.~1), and the geodesics are sensitive to that lumpiness.  We will investigate this question in two different contexts.  The first is that where we imagine a pair of ``satellites" (freely moving heavy masses), and we consider the distance between them.  The second context, which is free of any such extra matter dynamics, is that where we imagine an inflating spacetime whose spatial slices have the topology of the torus, $T^3$, and where we study curves with non-trivial homotopy on $T^3$.

In either case the basic expression for the distance is 
\beq\label{lineint}
{\cal S}(t) = \int_{X_1(t)}^{X_2(t)} \sqrt{h_{ij} dX^i dX^j}\ ,
\eeq
integrated along a curve $X^i(t,\sigma)$ in the spatial slice $t=$const.  In the satellite case, $X_I(t)$ are the locations of the satellites, and in the torus case we take $X_1(t)=X_2(t)$.  
The fluctuating spatial metric is given by \eqref{spacemet}.  It is convenient to group the metric perturbations together by defining
\beq
\Gamma_{ij}(x,t) = \gamma_{ij}(x,t) + 2\zeta(x,t)\delta_{ij}\ .
\eeq

Also, in some cases it is useful to define a {\it scale-dependent} metric, as introduced in \cite{Giddings:2011zd}, which only incorporates fluctuations above a certain wavelength.  To leading order in $H$, this can be defined by 
\beq\label{scalemet}
h_{ij}(q; x,t)=a^2(t) [\exp\{ \Gamma(q,x)\}]_{ij}, 
\eeq
 where $\Gamma_{ij}(q,t,x)$  is the scale-dependent metric fluctuation at scale $q$,
\beq\label{gammadefa}
\Gamma_{ij}(q,t,x) = \int^q_{L^{-1}} {d^3k\over (2\pi)^3} \left[ 2 \zeta(k; t,x)\delta_{ij} + \gamma_{ij}(k; t,x)\right]\ 
\eeq
and $\zeta(k; t,x)$, $\gamma_{ij}(k; t,x)$ are mode functions with comoving momentum $k$.\footnote{At higher orders, the limit in the integral \eqref{gammadefa} depends on $\Gamma$\cite{Giddings:2011zd}.} In what follows, we will assume $q<a(t)H(t)$, such that this expression will be constant in time to a very good approximation. In \cite{Giddings:2011zd}, a scale-dependent physical momentum was also introduced
\beq\label{scaledepp}
\kqi(k,x) = [e^{-\Gamma(q,x)/2}]_{ij} k_j \quad ; \quad \kq^2(k,x) = [e^{-\Gamma(q,x)}]_{ij} k_i k_j\ ,
\eeq
and it was then argued that leading IR-dependent higher-order effects are incorporated into the spectrum $P$ by writing the {\it tree-level} two-point function $P_0(k)$ instead as a function of $\kk(k,x)$: 
\beq\label{resum}
P(k,x)d^3k = P_0(\kk(k,x)){d^3\kk }\ .
\eeq

The line integral \eqref{lineint} can be expanded in terms of the metric fluctuation $\Gamma_{ij}$; if we use the scale-dependent metric \eqref{scalemet}, $\cals$ also depends on the scale $q$ determining which metric fluctuations are included.  For example, suppose that we choose comoving coordinates on an initial slice so that the satellite separation or torus cycle are in the $x$ direction.  In that case, we can compute the contribution of the fluctuations to the distance $\cals$ along a line of constant $y$ and $z$ connecting the endpoints; this can be expanded as 
\bea \label{W}
\cals &=& \int ds = a(t) \int dx \sqrt{\left[e^\Gamma\right]_{xx}}\nonumber \\
& =&a(t)\int dx(1+\frac{1}{2}\Gamma_{xx}+\frac{1}{4}\Gamma_{xi}\Gamma_{ix}-\frac{1}{8}\Gamma_{xx}\Gamma_{xx}+\dots)~.
\eea
Here, for brevity we suppressed the arguments of $\Gamma_{ij}$.
Next, we can compute the average effect of the fluctuations on this distance.  
In cases where our slices are chosen such that $\zeta=0$, then the linear contribution $\langle \Gamma\rangle$ vanishes,\footnote{Validity of such gauge fixing to second order is checked in the appendix.} and we need to consider quantities quadratic in $\Gamma$ to see an effect.  (While this no longer holds in the case of slow roll in a slicing with $\zeta\neq0$, later in the paper we will demonstrate slow-roll suppression of such contributions.)
One possibility is to look at the variance of $\cals$, via
\beq\label{stav}
\left\langle \cals  \cals\right\rangle\ .
\eeq
The contribution to this from the linear term in $\Gamma$ can be computed, but is found to be small (see section \ref{VLI}).  Intuitively, the linear contributions to $ \cals$ average out when integrated around the $x$-cycle.

The fluctuations also enter at quadratic order, suggesting a growing contribution to $\langle \cals\rangle$.  This can be written in terms of the variance of the fluctuations as
 \beq\label{W0}
 \left< \cals \right> = a(t)L\left(1+\frac{1}{2}\langle \zeta^2(x)\rangle +\frac{1}{4}\langle \gamma_{xi}(x)\gamma_{ix}(x)\rangle -\frac{1}{8}\langle \gamma^2_{xx}(x)\rangle+\dots\right)\ .
 \eeq
 Eq.~\eqref{Tensvar} and the analogous equation for $\zeta$ show that these contributions grow with time.  
 However, while the basic message of this equation -- growing contribution due to the variance -- is correct, this equation is not particularly physical as it stands.  The problem is that the curve $y=$const., $z=$const. is rather arbitrary from the viewpoint of the perturbed metric.  This has the effect of making the result \eqref{W} gauge-dependent, hence unphysical.  

Specifically, we could have described this computation by still working at constant $y$ and $z$, but in a different gauge from \eqref{transverse} and get a different result. The general gauge transformation corresponds to the action of the diffeomorphism $x^\mu\rightarrow x^{\mu\prime} = x^\mu - \epsilon^\mu(x)$ on the metric.  However, from the diffeomorphism symmetry of the basic expression \eqref{W}, we can alternately describe the situation by working with fixed metric, and consider a deformation of the {\it curve}; this corresponds to working in the $x'$ coordinates, and has the effect $x^\mu\rightarrow x^\mu+\epsilon^\mu$ on the curve.\footnote{What we have described is the standard switch between active and passive viewpoints, namely from a transformation on the fields to a transformation on the underlying coordinates.  Note also that in the case of the path connecting satellites, the diffeomorphism also acts on the satellite trajectories.}
 
To make a gauge-invariant statement, we would like a quantity independent of  such a variation, which we rewrite in general as 
 \beq\label{pathvar}
X^\mu(\sigma) = X^\mu_0(\sigma)+\delta X^\mu(\sigma) \ ,
\eeq
where $X_0^\mu(\sigma)$ is the original path. In our particular example 
\beq\label{xpath}
X^\mu_0=(0,x,0,0)\ .  
\eeq
We will first focus on the purely spatial gauge transformations $\delta X^i$, viewing $\epsilon^0$ as fixed by choice of physical ``clock," determined by either of the conditions \eqref{constphi} or \eqref{zerozeta}; we will examine these in particular cases.
While $\delta X^i$ introduces arbitrariness into $\cals$, one might expect that there is a unique choice, corresponding to a geodesic in the new metric.  Then, one could calculate the average variation of the length of this geodesic, as a gauge invariant.  
 
In order to explore this, we consider the change of $\cals$ under a combined variation $\delta X^i$ and 
\beq\label{metvar}
h_{ij} =  h_{0,ij} +\delta h_{ij}
\eeq
where $h_{0,ij}$ is given in \eqref{classsoln} (we work at fixed time).
Expanding to second order, this change takes the form
\bea\label{varia}
\cals[X_0+\delta X, h_0+\delta h] &=& \cals[X_0,h_0] +\frac{\delta \cals}{\delta h}\delta h +\frac{\delta \cals}{\delta X}\delta X \nonumber\\
& +&\frac{1}{2}\frac{\delta^2 \cals}{\delta h^2}\delta h^2+\frac{1}{2}\frac{\delta^2 \cals}{\delta X^2}\delta X^2+\frac{\delta^2 \cals}{\delta X\delta h}\delta X\delta h +\dots\ ,
\eea
where integrals are implicit and all variational derivatives  are evaluated at $X_0$, $h_0$.
The linear term in $\delta X$ vanishes because  the original curve was a geodesic.  The linear and second-order terms in $\delta h$ combined are given in terms of $\gamma_{ij}$ and $\zeta$ by \eqref{W}.   Our problem, therefore, is to understand the remaining $\delta X$-dependent terms. 

Following the preceding discussion, we can try to fix $\delta X$ by the condition that the perturbed curve is a geodesic in the perturbed metric.  Working to linear order in both perturbations, this is the condition
\beq\label{geodv}
0=  {\delta\cals\over \delta X}\Bigl |_{\genfrac{}{}{0pt}{}{X_0+\delta X,}{h_0+\delta h ~}} = \frac{\delta^2 \cals}{\delta X^2}\delta X^2+\frac{\delta^2 \cals}{\delta X\delta h}\delta X\delta h\ ,
\eeq
with second derivatives evaluated at $X_0, h_0$.  This can be rewritten 
\beq\label{geodv2}
\delta X {\delta\over \delta X}{D_0^2  X\over ds_0^2} = 	- \delta h{\delta\over \delta h} {D^2X_0\over ds^2} \ ,
\eeq
found by varying the usual geodesic equation. For a solution of \eqref{geodv}, \eqref{geodv2}, the last two terms of \eqref{varia} may then be written in terms of 
\beq\label{transvar}
\frac{\delta^2 \cals}{\delta X^2}\delta X^2 = \int ds {d\delta X^i\over ds} {d\delta X^j\over ds}\left(h_{ij}-  h_{ik} {dX^k_0\over ds}h_{jl} {dX^l_0\over ds}\right)\ .
\eeq

For our x-path \eqref{xpath}, the geodesic equation becomes
\beq
\p_x^ 2\delta X^\alpha=-\Gamma^\alpha_{xx}=a^{-2}(t)\left(-\p_x\delta h_{x\alpha}+\frac{1}{2}\p_\alpha\delta h_{xx}\right)~,\qquad \alpha=y,z\ , \label{yzeqn}
\eeq
where in the latter equation we find the linearized Christoffel symbols.  A deformation $\delta X^x$ corresponds to a reparameterization of the path and is unconstrained.  We will study the problem of solving these equations both for $T^3$ and for the satellites, and of determining the resulting change in path length in \eqref{varia}.

\section{Cycles on the torus}

\subsection{Quantization on the torus}

For simplicity we begin with the case of the spatial manifold $T^3$, and with no matter present.   We take the background metric to be \eqref{classsoln}, with the scale factor of dS, \eqref{dsscale}, now with the identifications $x^i\sim x^i+L$. 
The preceding discussion of fluctuations and gauge fixing thus carries over with minor modification.\footnote{For previous discussion of quantization on $T^3$, see \cite{Tsamis:1993ub}.}  

To investigate the possibility of residual gauge transformations, consider the positive definite integral
\beq
0\leq \int d^3x \left[(\partial_i \epsilon_j)(\partial_i \epsilon_j) + {1\over 3} (\partial_i\epsilon_i)(\partial_j\epsilon_j)\right] = -\int d^3x \epsilon_i (\nabla^2 \epsilon_i + {1\over 3} \partial_i\partial_j \epsilon_j)\ 
\eeq
where the last equality arises from integration by parts, with no surface term for single-valued $\epsilon^i$.  For a residual gauge transform, the latter integral vanishes by \eqref{residxm}, and then by positive definiteness we find that $\partial_i \epsilon_j=0$.  Thus, the only residual gauge transformation is a trivial shift by a constant.

In the transverse-traceless gauge \eqref{transverse}, \eqref{zerozeta}, the quadratic action is simply that for tensors, \eqref{tensact}.  
The corresponding general linearized solution is given by the mode expansion, 
\begin{equation}\label{gravdecomp}
    \gamma_{ij}(x) = \sum_{I}{\ep}^I_{ij}\gamma_{0I}(t) +\sum_{s=+,\times} \sum_{{\bf k}\neq 0}\frac{1}{L^ 3}\left[
        b^{s}_{\vect{k}}\ep^s_{ij}(\vect{k})\gamma_k(t)
        + b^{s\dagger }_{-\vect{k}}\ep_{ij}^{s*}(-\vect{k})\gamma^{*}_k(t)\right]
        e^{i\vect{k}\cdot\vect{x}}~.
\end{equation}
Here the first term is a space-independent zero mode contribution with mode index $I$.  The momentum sum is over quantized values $\vect{k} = (2\pi/L)(n_x, n_y, n_z)$, with integer $n_i$; specifically, quantization is related to that in infinite space by the replacements
\beq
\int {d^3k\over (2\pi)^3}\rightarrow {1\over L^3}\sum_{\vect{k}}\quad ;\quad (2\pi)^3\delta^3\left(\vect{k}-\vect{k'}\right)\rightarrow L^3 \delta_{\vect{k}\vect{k'}}\ .
\eeq
The ladder operators $b^{s}_{\vect{k}}$, with  $s$ denoting the helicity, therefore satisfy the commutation relations
\beq
\left[ b^{s}_{\vect{k}},  b^{s\dagger }_{\vect{k}'}\right] =L^3\delta_{ss'}\delta_{\vect{k}\vect{k}'}\ .
\eeq
The polarization tensors
$\ep^s_{ij}$ are chosen to satisfy
the transversality and tracelessness conditions
$\ep^s_{ii}(\vect k)=k_i\ep^s_{ij}(\vect k)=0$,
together with a completeness relation obtained by tracing over
spatial indices, $\ep_{ij}^s(\vect{k})\ep_{ij}^{\ast s'}(\vect{k})=2\delta_{ss'}$.  The mode functions are the same as those for a massless scalar, up to normalization; they are most easily written in terms of the conformal time
\beq
\eta = -{1\over Ha(t)} 
\eeq
as
\beq\label{modefcns}
 \gamma_k (\eta) = 
    \frac{H}{\sqrt{k^3}}(1+ik\eta)e^{-ik\eta}\ .
 \eeq
 
The quantization of the zero modes is handled separately.  In the case of finite volume, these modes have finite norm.
The equation of motion for them has spatially homogenous solutions of the form
\beq
\gamma_{0I} = Q_I + P_I\int \frac{dt}{a^3} 
\eeq
with conjugate variable
\beq
\pi_{0I} = a^3\dot\gamma_{0I} = P_I~.
\eeq
In particular, the canonical commutation relations for $\gamma$, together with completeness, imply commutation relations of the form
\beq
[Q,P]=\frac{i}{V}
\eeq
where $V$ is the volume of the compact space, and we drop the mode indices.  The ``dS-invariant" zero-momentum vacuum $P\left|P=0\right>=0$ is non-normalizable just like the ground state for an ordinary free particle \cite{Kirsten:1993ug}; the corresponding wavefunction is $\gamma_0$ independent, so it gives a completely indefinite $\gamma_0$.

To specify an initial size/shape for the torus, we instead choose the vacuum wave function for the zero modes $\gamma_0$ to be gaussian wave packets peaked around some definite values (here taken to be zero) in position space
\beq\label{wavefunct}
\Psi(Q) = \left(\frac{1}{2\pi(\Delta Q)^2}\right)^{1/4}\exp\left[-\frac{Q^2}{4(\Delta Q)^2}\right]~.
\eeq
The initial velocity due to the finite width of the wave function quickly redshift away, and since
the energy density of the zero mode redshifts away like $a^3 \dot\gamma_0^2 \propto 1/a^3$ it can effectively be ignored.

This type of initial state belongs to a class of $O(4)$ invariant Hadamard Fock vacuum states \cite{Allen:1987tz}, which can be constructed by defining 
\bea
P&=&\left( A a_0 +A^*a_0^{\dagger}\right)\nonumber\\
Q&=&\left( B a_0 +B^*a_0^{\dagger}\right)
\eea 
 where the commuation relation, $\left[a_0,a_0^{\dagger}\right]=1$, implies
 \beq
 A^*B-B^*A =\frac{i}{V}~.
 \eeq
 The Hadamard Fock vacua are then defined by
 \beq
 a_0 \left| 0\right>_{HF} = 0~.
 \eeq
 With the choice $A = -i/(\sqrt{2} V\Delta Q)$ and $B= \Delta Q/\sqrt{2}$, the vacuum wave function in position space reduces exactly to the one in eq.~(\ref{wavefunct}); see \cite{Allen:1987tz} .
 
 \subsection{Growth of variance}
 
 The two-point function gives an important measure of the gravitational fluctuations; specifically, as in  \eqref{Tensvar}, consider the double trace,
 \beq
\langle \gamma^2(x,x')\rangle = \frac{1}{4}\langle \gamma_{ij}(x) \gamma_{ij}(x')\rangle =\langle \gamma_0^2(t)\rangle+ {1\over L^3}\sum_{\vect{k}}{H^2\over k^3}(1+k^2\eta^2)e^{i\vect{k}\cdot(\vect{x}-\vect{x'})}  \ .
\eeq
Here, 
the IR divergence of the analogous  expression for non-compact space has been regulated by the finite size of the torus, which plays the role of a comoving IR cutoff, like was used in \cite{Giddings:2010nc,Giddings:2011zd,Brandenberger}.  The zero-mode contribution asymptotes to a constant $\propto (\Delta Q)^2$ at long time.
The coincident limit $\vect{x}=\vect{x'}$, which gives the variance, is UV divergent, but may be regulated by choosing a minimum physical separation $a(t)|\vect{x}-\vect{x'}|\sim 1/H$, effectively providing a UV cutoff at $k\approx 2\pi a(t)H$.  This introduces a logarithmic time-dependence into the variance, as in the non-compact case; specifically, 
\beq\label{vargrowth}
\langle \gamma^2(x)\rangle \approx 2\Hotps\int_{2\pi/L}^{2\pi a(t)H} \frac{dk}{k}(1+k^2\eta^2)\approx 2\frac{H^ 2}{(2\pi)^ 2}\log[a(t)HL]\ ,
\eeq
which grows linearly with time.  
We will also need the variance of specific tensor components, which can be derived using the  sum over polarizations\cite{Weinberg:2008zzc}
\bea\label{polarsum}
\omega_{ij,kl}(\vect q)=\sum_s \ep^s_{ij}(\vect q)\ep^{s*}_{kl}(\vect q) &=& \delta_{ik}\delta_{jl}+ \delta_{il}\delta_{jk}-\delta_{ij}\delta_{kl}\\
&+&\delta_{ij}\khat_k\khat_l + \delta_{kl}\khat_i\khat_j- \delta_{ik}\khat_j\khat_l - \delta_{il}\khat_j\khat_k-\delta_{jk}\khat_i\khat_l-\delta_{jl}\khat_i\khat_k + \khat_i\khat_j\khat_k\khat_l\nonumber
\eea
where $\vect {\khat}$ is the unit vector in the direction $\vect{q}$.  For an arbitrary unit vector $\vect{n}$, we find\cite{Giddings:2010nc}
\beq\label{tensvar}
n_i n_j \langle \gamma_{il}\gamma_{lj}\rangle\approx \frac{4}{3} \langle \gamma^2(x)\rangle\quad ,\quad n_in_jn_kn_l\langle \gamma_{ij}\gamma_{kl}\rangle\approx \frac{8}{15} \langle \gamma^2(x)\rangle\ .
\eeq

We can use  \eqref{tensvar} to evaluate the tensor variance in \eqref{W0}; for generality we likewise include the scalar contribution, although in the present section it is gauged to zero.
The result is that on the unperturbed path the line integral yields
 \beq\label{W02}
 \left< \cals \right> = a(t)L\left(1+\frac{1}{2}\langle \zeta^2(x)\rangle +\frac{4}{15}\langle \gamma^2(x)\rangle +\dots\right)\ .
 \eeq

In de Sitter space or its $T^3$ version, the variance, $\langle \gamma^2(x)\rangle$, keeps growing linearly with time like $H^3t$ at leading order, and the effect becomes order one at a time scale of order $t  \sim R_{dS}S_{dS}$.  (This is in parallel with the apparently related case of black hole evolution \cite{Giddings:2007ie, Nimatalk,Giddings:2010nc}.) But even at the end of  the non-eternal regime of chaotic inflation, we have $\langle \gamma^2(x)\rangle \sim 1$ at the time of reheating, if initial conditions are set at the end of the self-reproduction regime.

\subsection{The line integral on the torus}

\begin{figure}[!hbtp] \label{fig2}
\begin{center}
\includegraphics[width=12cm]{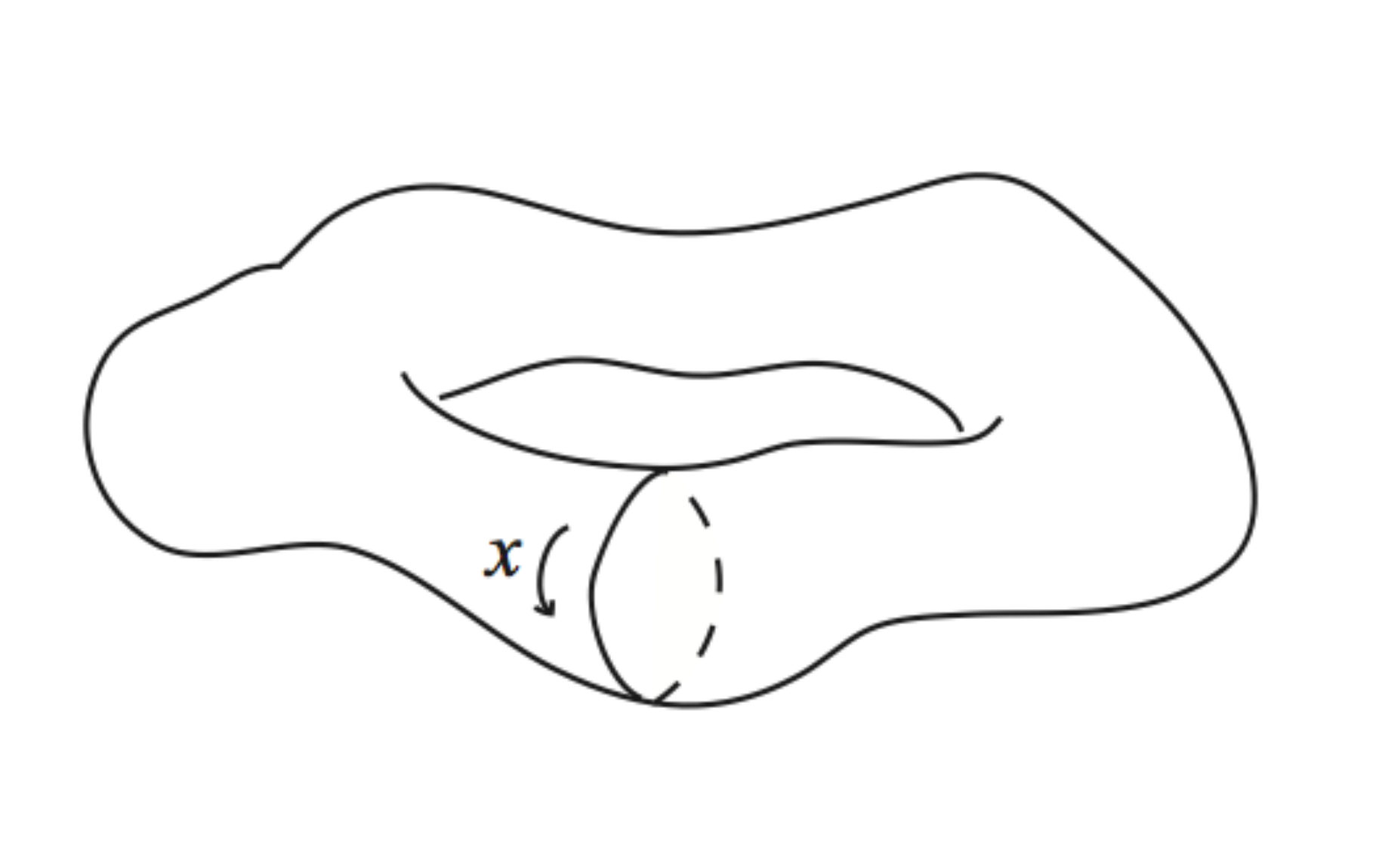}
\end{center}
\caption{Pictured is a torus, made ``lumpy" by metric perturbations, along with a representative cycle along which distance $\cals$ can be calculated.}
\end{figure}

As argued in section 4, the line integral along the unperturbed path does not give gauge-invariant information, and it is no longer a geodesic in the perturbed metric.  We can try to remedy these by trying to solve the geodesic equation in the constant-time hypersurface to second order, in order to find the curve with the shortest distance around the torus.

The geodesic equation \eqref{yzeqn} can be integrated, giving 
\beq\label{firstint}
\partial_x \delta X^\alpha = a^\alpha -\int_0^x dx' \Gamma^\alpha_{xx}\ ,
\eeq 
with integration constants $a^\alpha$.  However, for a closed geodesic, \eqref{firstint} should be periodic, which is the condition  
\beq
\int_0^L dx \Gamma^\alpha_{xx} =0\ .
\eeq 
From the expression \eqref{yzeqn} we find that this requires
\beq
\partial_\alpha \int_0^L dx\, \delta h_{xx} =0\ .
\eeq

Let us first consider the case with vanishing fluctuations  $\delta h_{xx}=0$.  Then a perturbed geodesic can be found by integrating and enforcing periodicity.  It is useful to introduce a notation for an average over the curve,
\beq\label{avfluct}
[f] = {1\over L} \oint dx f(x)\ ,
\eeq
with $\oint$ the integral from 0 to $L$, 
in terms of which we find 
\beq\label{nearg}
\delta X^\alpha = \delta X^\alpha_1\equiv  -{1\over a^2(t)} \int_0^x dx' ({\delta h_{x\alpha} - [\delta h_{x\alpha}]}) ,
\eeq 
up to an overall constant shift.  We can then evaluate the total effect on $\langle \cals \rangle$; since the geodesic equation is satisfied and can be written as  \eqref{geodv},
we find that the last two terms of \eqref{varia} have average
\beq\label{WX2}
\frac{1}{2}\left<\frac{\delta^2 \cals}{\delta X^2}\delta X^2\right>+\left<\frac{\delta^2 \cals}{\delta X\delta h}\delta X\delta h\right> = -{1\over 2}  \left<\frac{\delta^2 \cals}{\delta X^2}\delta X^2\right>=-{a(t)\over 2}\oint dx \left<\p_x\delta X_\alpha\p_x\delta X_\alpha\right>\ 
\eeq
({\it c.f} \eqref{transvar}). Using \eqref{nearg} then gives 
\beq\label{X2}
{1\over 2 }  \left<\frac{\delta^2 \cals}{\delta X^2}\delta X^2\right> = {1\over 2 a^3(t)}  \oint dx\left\langle\delta h_{x\alpha} \delta h_{x\alpha} - [\delta h_{x\alpha}][\delta h_{x\alpha}]\right\rangle \ .
\eeq 
The last term is subdominant, due to cancellation of the fluctuations in the path average, as in \eqref{stav}.

Next, let us find the additional terms present when $\delta h_{xx}\neq 0$.  First, there are the quadratic terms in $\delta h$ already present in \eqref{W0}.  Then, there is also a cross term linear in 
$\delta X$ and $\delta h$, 
\beq
{1\over 2 a(t)}\oint dx \partial_\alpha\delta h_{xx} \delta X^\alpha\ .
\eeq
If we write $\delta X = \delta X_1 +\delta X_2$, with $\delta X_1$ given in \eqref{nearg},  
then the term involving $\delta X_1$ averages to zero
since $\langle \gamma_{xx}\gamma_{x\alpha}\rangle = 0$ for $\alpha\neq x$.  There are also quadratic terms in $\delta X$.  If $\delta X_1$ and $\delta X_2$ are uncorrelated, $\langle \partial_x \delta X^\alpha_1 \partial_x \delta X^\alpha_2\rangle=0$, so we have a term as in \eqref{WX2} involving only $\delta X_2$.

Thus, combining these terms, 
the first, leading, term of \eqref{X2} and the other terms in \eqref{varia} gives
\bea
\left< \cals \right> &=& a(t)L\left[1 - \frac{1}{L}\oint dx\left(\frac{1}{4}\left<\gamma_{xi}\gamma_{ix}\right>- \frac{3}{8}\left<\gamma_{xx}\gamma_{xx}\right> \right)  +\calo\left(\left<\gamma^2(x)\right>^2\right)\right]\cr
&&\quad+{1\over 2a(t)} \left\langle\oint dx  \partial_\alpha \delta h_{xx}\delta {X}^\alpha_2\right\rangle+ {a(t)\over 2}\oint dx \left<\p_x\delta X^\alpha_2\p_x\delta X^\alpha_2\right>
\ .
\eea
Consider the case $\delta X_2=0$.  Using eq.~\eqref{tensvar}, we find
\beq\label{Wred}
\left< \cals \right> = a(t)L\left[1-\frac{2}{15}\left<\gamma^ 2(x)\right>  +\calo\left(\left<\gamma^2(x)\right>^2\right)\right]    
\eeq
in the spatially-flat time slicing \eqref{zerozeta}, with $\langle\gamma^2\rangle$ as in \eqref{vargrowth}.
Comparing with eq.~\eqref{W02}, we observe that, while the circumference around the original curve becomes {\it larger} due to the growing variance,  {\it we can always find a nearby curve that is shorter than the original curve, by an amount that grows with the variance of the metric}. 

For general $\delta h_{xx}$ there will not be a  neighboring geodesic, but the length of the curve can be reduced further by choice of $\delta X_2$.  For example, with constant $\delta X_2$, 
$\cals$ changes by
\beq
{1\over 2a(t)} \delta {X}^\alpha_2\oint dx \partial_\alpha \delta h_{xx} \ .
\eeq
So, when $\partial_\alpha \oint dx\delta h_{xx}\neq 0$, $\langle \cals\rangle$ is further reduced by $\delta X_2$ with a particular correlated sign.

An intuitive picture for our statements is the following.  First, variations $\delta h_{xy}$, $\delta h_{xz}$ correspond to a contraction of space in some directions, and  expansion in other directions.  The basic idea is that by choice of the path, we can take advantage of the contraction and shorten the total path length.  On the other hand, consider a variation $\delta h_{xx}(y,z)$.  This has the effect of changing the circumference of the $x$-cycle in a $y$ and $z$ dependent fashion, as sketched in fig.~2.  We can decrease the length of the curve by sliding it towards smaller circumference, although reaching a minimum of the circumference, corresponding to a geodesic, may require a large deformation.

But, since the length can always be lowered below \eqref{Wred}, we have an invariant statement: for some curves in the relevant homotopy class, the average $\langle \cals\rangle$ is reduced by at least the amount in \eqref{Wred}, and this reduction grows with the variance of $\gamma$.

\section{Satellite distances and correlator modifications}\label{SDCM}

An alternative way to make diffeomorphism invariant statements is to consider a length between two physical comoving particles in an inflationary spacetime; in the classical limit we might also describe them as ``satellites."  We assume that at some initial time they are separated by comoving distance $L$, in the $x$ direction.  

If we compute the distance between the satellites as a function of time, a number of effects besides the inflationary expansion can contribute.  First, the satellites can have an initial velocity.  Classically, this may be set to zero, but in the quantum case it has nonzero spread due to position/momentum uncertainty.  Secondly, with zero such velocity the satellites in the zeroth-order metric stay at fixed comoving location in the $(x,y,z)$ coordinates, but metric fluctuations drive motion of the satellites.  Thirdly, at sufficiently long times new ``Boltzmann" satellites that are indistinguishable from the original satellites can be created, confounding the definition of the observables\cite{GiMa}.  Finally, we have contributions of the metric fluctuations to the path length between the satellites, both due to local changes in the length, and due to the need to perturb to a new path satisfying the geodesic equation in the new metric on a constant-time slice, as described in the preceding sections.

As we will argue in subsection \ref{sublead}, the inflationary expansion rapidly redshifts away an initial velocity, as well as gradients responsible for satellite motion, and so these effects, as well as those of Boltzmann satellites, are small on timescales where metric fluctuations make significant contributions to the path length.  For that reason, we will first consider the latter set of effects.

\subsection{Perturbations in path length}\label{PiPL}

We begin by investigating the change in length of the path between two satellites with fixed comoving coordinates.  
Thus, we want the contributions of metric and path fluctuations to the integral \eqref{lineint}, with endpoints taken to lie at 
$X_1(t)=(0,0,0)$ and $X_2(t)=(L,0,0)$.  In the case of slow-roll with the comoving gauge \eqref{constphi}, there is a contribution due to $\langle \zeta\rangle\neq0$; however, we will see in the next subsection that this is small relative to the other effects we consider.  To second-order in fluctuations of the metric and path, the change in length is  
again given by \eqref{varia}.  In the satellite case, the geodesic equation \eqref{geodv} can be solved for the geodesic connecting the endpoints.  This allows us to eliminate the cross term proportional to $\delta h \delta X$ in \eqref{varia}.  The result for the average $\langle\cals\rangle$ is
 \beq\label{expD00}
\left<\cals\right> =\cals_0
+ a(t)\int_0^L dx\left(\frac{1}{4}\left<\Gamma_{xi}\Gamma_{ix}\right>-\frac{1}{8}\left<\Gamma_{xx}\Gamma_{xx}\right>\right)-{a(t)\over 2}\int_0^L dx \left<\p_x\delta X_\alpha\p_x\delta X_\alpha\right>+\cdots~.
\eeq 
In order to evaluate the last term in this equation, we must solve the geodesic equation, which to linear order in perturbations is given by  \eqref{yzeqn}.
The solution is of the form
\beq
\delta X^\alpha =-\int_0^x dx'\int_0^{x'}dx'' \Gamma_{xx}^\alpha + a^\alpha x +b^\alpha
\eeq
where again $\alpha=(y,z)$.
The boundary condition at $x=0$ gives $b^\alpha=0$, and the boundary condition at $x=L$ fixes $a^\alpha$.  If we introduce the notation 
\beq\label{fdef}
f^\alpha = -\int_0^x dx' \Gamma_{xx}^\alpha
\eeq 
and use the path-average notation \eqref{avfluct}, the solution becomes 
\beq
\delta X^\alpha = \int_0^x dx'(f^\alpha - [f^\alpha])\ .
\eeq 
Thus, the second-order variation in the path length, \eqref{expD00}, is fully determined in terms of the two-point functions of the metric fluctuations.

The fact that $\zeta$ and $\gamma$ are independent random variables at quadratic level is sufficient for us to be able to minimize the path in $\zeta$ and $\gamma$ independently. Let us therefore consider the physical time-slicing given by $\phi=\phi_0$, \eqref{constphi}, and first focus on the 
effect of $\zeta$; $\gamma$ fluctuations are subdominant in the slow-roll expansion. In this case we have leading contribution $\left<\Gamma_{xi}\Gamma_{ix}\right>= \left<\Gamma_{xx}\Gamma_{xx}\right> = 4\left<\zeta^2\right>$, thus
\beq\label{expD01}
\left<\cals\right> =\cals_0
+ \frac{1}{2}a(t)\int_0^L dx\left<\zeta^2\right>-{a(t)\over 2}\int_0^L dx \left<\p_x\delta X_\alpha\p_x\delta X_\alpha\right>~.
\eeq

Let us first consider the situation where we neglect the perturbation of the path, as was done in \cite{Urakawa:2010it,Urakawa:2010kr,Urakawa:2011fg} (see {\it e.g.} sect. IV.B. of \cite{Urakawa:2010it} or eqs. (66), (67) of \cite{Urakawa:2010kr}) and \cite{Gerstenlauer:2011ti}.   Then we simply have 
\beq\label{D1}
\left<\cals\right> =a(t)L\left(1+
 \frac{1}{2}\left<\zeta^2\right>+\dots \right)\approx aLe^{\frac{1}{2}\left<\zeta^2\right>}
\eeq
As described in section \ref{tensfluc}, we might consider the two-point correlation function, as a function of path length.  In terms of coordinate separation, this correlation function is 
\beq
\left<\zeta(0)\zeta({\bf x})\right> = \int \frac{d^3 k}{(2\pi)^3}\frac{H^2}{4\epsilon k^3}\left(\frac{k}{aH}\right)^{n_s-1}e^{i{\bf k}\cdot{\bf x}}~.
\eeq 
If we take ${\bf x}= (L,0,0)$, and write the correlation function in terms of  $\left<\cals\right>$ instead, then we obtain 
\beq\label{pdistcorr}
\left<\zeta(0)\zeta(x(\langle\cals\rangle)\right> = \int \frac{d^3 \tilde k}{(2\pi)^3}\frac{H^2}{4\epsilon \tilde k^3}\left(\frac{\tilde k}{aH}\right)^{n_s-1}e^{\frac{1}{2}(n_s-1)\left<\zeta^2\right>}e^{i\tilde  k_x \langle\cals\rangle/a}~.
\eeq 
where we defined $\tilde k\equiv k e^{-\frac{1}{2}\left<\zeta^2\right>}$. In \cite{Urakawa:2010it,Gerstenlauer:2011ti} this object was considered as a candidate IR-safe observable, although this construction is not necessarily relevant for a late-time observer of the CMB sky for which an IR-safe observable was instead constructed in \cite{Giddings:2011zd}. One problem with the construction above from the point of view of a late-time observer is that the curvature perturbation is not conserved on super-horizon scales when written in terms of the geodesic distance, since the geodesic distance receives large corrections on large scales. This implies that in terms of the of the average of the actual distance $\langle\cals\rangle$ along the original path on the reheating surface, neglecting the perturbation of the path, the spectrum becomes\footnote{The same formula was given by \cite{Senatore}.} 
\beq
P_{\zeta}(\tilde k) = P^{(0)}_{\zeta}(\tilde k)\left[1+\frac{1}{2}(n_s-1)\left<\zeta^2\right>+\dots\right]~,
\eeq
where only modes that exit the horizon after the mode $\tilde k$, but before reheating, contributes to the variance. Thus, for a cosmologically relevant scale exiting the horizon only $60$ e-folds before the end of inflation this effect is order $H^2N$, which is small. But for very large scales, which exited the horizon $N\sim 1/H^2$ e-folds (or $\Delta t\sim RS$) before the end of inflation, the effect is large.\footnote{Note that this is different than the effect studied in \cite{Giddings:2010nc}. There new semi-classical relations was used to find the effect of long wavelength modes on the correlation functions written in comoving coordinates (not in terms of geodesic distance on the reheating surface). In comoving coordinates the curvature perturbation is conserved on super horizon scales, but long wavelength modes, exiting the horizon before the observed mode, shift the background of the observed mode when it exits the horizon as in (\ref{resum}). To a late-time observer, comparing correlation functions at different scales, this becomes a physically observable effect \cite{Giddings:2011zd}.}

Moreover, we have argued that the original path does not have any invariant meaning in the perturbed metric; the natural choice is to perturb to the new path which is a geodesic in the perturbed metric.  We will again find this has a significant effect.

Specifically, again focus just on the dominant scalar fluctuations, with
\beq
\Gamma_{xx}^\alpha = -\p_\alpha\zeta\ .
\eeq
Using the notation \eqref{fdef}, we now have
\beq
f_\alpha(x) = \int_0^x dx' \p_\alpha \zeta(x')
\eeq
Thus, the contribution in \eqref{expD01} from the path perturbation becomes
\beq
{a(t)\over 2}\int_0^L dx \left<\p_x\delta X_\alpha\p_x\delta X_\alpha\right> ={a(t)\over 2} \int_0^L dx\left<\left(f_\alpha(x)-[f_\alpha]\right)^2\right>~.
\eeq
We then insert the mode expansion for $\zeta$, and obtain
\beq
\left<\left(f_\alpha(x)-[f_\alpha]\right)^2\right> = \int\frac{d^3 k}{(2\pi)^3}\frac{k_\alpha^2}{k_x^2}\left| e^{ik_x x} +\frac{1}{iL k_x}\left(1-e^{ik_x L}\right)\right|^2\left<\zeta_{\bf k}\zeta_{-{\bf k}}\right>~.
\eeq
Next, we integrate over $x$, and define a variable $r=k_x/k$, so that $k_\alpha^2 = k^2 (1-r^2)$.  These steps give
\beq\label{intexp}
\int_0^L dx\left<\left(f_\alpha(x)-[f_\alpha]\right)^2\right>= \int {k^2 dk\over (2\pi)^2} \left<\zeta_{\bf k}\zeta_{-{\bf k}}\right> \int_0^1 dr {1-r^2\over r^2} {(kLr)^2 + 2\cos(kLr) -2\over k^2r^2L}\ .
\eeq

At large $k$, the integral over $r$ in \eqref{intexp} behaves as $\sim \pi k L^2/6$.  One can readily see that the dominant contribution to this behavior comes from the regime of small $r$, that is $k_x\ll k$, $|k_\alpha|\sim k$.  For spectral index $n_s\geq0$, this produces a UV divergence in \eqref{intexp}.  For example, in the scale-invariant case $n_s=1$, with $\left<\zeta_{\bf k}\zeta_{-{\bf k}}\right> = A[1+(k/aH)^2]/k^3$, and introducing a comoving UV cutoff $\Lambda_{UV}$,
the contribution to the path length becomes 
\beq
{a(t)\over 2} \int_0^L dx\left<\left(f_\alpha(x)-[f_\alpha]\right)^2\right> = a(t) L\left [\frac{A}{48\pi} \left(\Lambda_{UV} +\frac{\Lambda_{UV}^3}{3(aH)^2}\right)L+\dots\right]
\eeq 
where the dots represent terms that grow less rapidly with the cutoff.  

Note that working with the scale-dependent metric \eqref{gammadefa} with scale $q$ corresponds to taking $\Lambda_{UV}=q$.  If we wish to account for all the metric fluctuations that are outside the horizon scale at a given time $t$, we should take $\Lambda_{UV} = a(t) H$, corresponding to a cutoff on physical momenta $\sim H$.  In this case, we find that the corrected formula for the path length takes the form 
\beq
\left<\cals\right>= a(t)L \left(1- \frac{A}{36\pi} a(t) H L + \dots\right)
\eeq 
For scalar fluctuations, $A=H^2/4\epsilon$; similar contributions are obtained from tensor fluctuations, with $A\sim H^2$.  We find that the corrections to the path length due to the fluctuations rapidly become comparable to the length of the unperturbed path.  For slow-roll inflation, this happens on a rapid time scale $t\sim H^{-1} \log[\epsilon/(H^3 L)]$, or when the satellite separation is of size
\beq
a(t)L\sim \epsilon H^{-3}
\eeq
(the result just accounting for tensor fluctuations is similar, without $\epsilon$).
This effect completely overwhelms that of eq.~(\ref{D1}), considered in \cite{Urakawa:2010it,Urakawa:2010kr,Urakawa:2011fg}; the basic picture is that the curve can significantly shorten its length by taking advantage of the short-wavelength variations in the scale factor. While these effects might be removed by sufficient coarse-graining, taking the scale $q\sim 1/L$, we conclude that there are 
large fluctuations in the actual geodesic distances between satellites at these scales. These provide an apparent obstacle to a characterization of the geometry in terms of lengths of its geodesics, at distance scales longer than $\sim 1/H^3$.  
And,  if correlators are defined directly in terms of the proper distance in the three-geometry, as in \eqref{corrFL} or eq.~\eqref{pdistcorr}, this large fluctuating contribution likewise becomes significant.\footnote{One might anticipate similar effects for the case of curves with constant acceleration, noted above, but checking this is left for future work.}

\subsection{Subleading effects}
\label{sublead}

This subsection argues that other contributions to the satellite separation are subleading to the effect in the preceding subsection; readers willing to take that for granted may wish to skip to the next section.  There are multiple possible sources for such effects.  The first category involves satellite dynamics:  the possibility of an initial velocity for the satellites, or, in the quantum context, spreading of the wavepacket or even production of new satellites.  The second category arises from metric fluctuations: satellites may move in response to metric perturbations, or there can be corrections to their separation from a one-loop contribution to $\langle\zeta\rangle$.  We consider these in turn.

\subsubsection{Satellite dynamics}\label{satdyn}

First consider the case where a satellite has an initial velocity in the comoving coordinates, $v=dx^1/d\tau$.  The Killing vector \eqref{kvect} then implies that
\beq
k\cdot {dx\over d\tau} = a^2(t) v
\eeq
is conserved.  Thus, any initial comoving velocity redshifts away as $1/a^2(t)$, and the satellite motion is rapidly dominated by the expansion.

In a quantum framework, one should of course consider a wavepacket describing the satellite motion.  Such a wavepacket is given by
\beq
\psi(x) = \int {d^3k\over (2\pi)^3} f(\bfk) u_k(t) e^{i \bfk\cdot\bfx}\ ,
\eeq
for some initial momentum-space profile $f(\bfk)$.  Here the temporal wavefunctions are solutions of 
\beq
{1\over a} {d\over dt} \left(a^3 {du_k\over dt}\right) + (k^2 + m^2 a^2)u_k=0\ .
\eeq
For dS, these take the form
\beq
u_k(t)= \eta^{3/2} H_\nu^{(1,2)}(k\eta)
\eeq
in terms of the conformal time \eqref{conftime}, and with $\nu^2 = 9/4 - m^2/H^2$.  The positive frequency solution is given by $H^{(2)}$.  On time scales short as compared to $1/H$, we have $|k\eta|\gg1$, and an initial wavepacket with width $\Delta x\sim 1/\Delta k \ll 1/H$ can be easily seen to evolve as in flat space, and in particular spreads as $\Delta x(t)\sim t \Delta k/m$. At longer time scales, $k\eta\ll1$, and one may use the asymptotics
\beq
H^{(2)}_\nu\sim {i\over \pi} \Gamma(\nu) \left({2\over k\eta}\right)^\nu 
\eeq
to find
\beq\label{superH}
\psi(x)\sim \eta^{3/2-\nu} \int {d^3k\over (2\pi)^3} f(\bfk) {1\over k^\nu} e^{i \bfk\cdot\bfx}\ ,
\eeq
up to an overall constant.  Thus, the spread $\Delta x\sim 1/\Delta k$ in comoving coordinates $x$ is frozen in time.\footnote{For $m\gg H$, the factor in front of the integral in \eqref{superH} takes the form $\eta^{3/2-\nu} \sim e^{-3Ht/2 -imt}$, corresponding to Hubble dilution with massive oscillation.}  This is consistent with the classical result that any initial velocity is rapidly redshifted to zero, having little subsequent effect.

Finally, if satellites are treated as quantum objects, they may be created by fluctuations, just as with elementary particles, Boltzmann brains, or any other object in an inflating universe.  If the number of microstates of such a satellite is $\cal N$, the production probability per unit four-volume is of size $\exp\{-\cal N\}$.  Then, after $N$ efolds, the total number produced is $\sim \exp\{N-{\cal N}\}$. Since $\cal N$ is bounded by the largest black hole in dS, and thus ${\cal N}\lesssim S$, we find a number of ``Boltzmann" satellites comparable to the original number of satellites by $N\sim S$ efolds.  Before $N\sim \cal N$, this is a small effect.

\subsubsection{Force on satellites}

Satellites can also deviate from fixed comoving position due to forces exerted by the metric perturbations.  To check that this effect is also small, first note that the satellite positions will be fixed in a different gauge, corresponding to geodesic-normal coordinates:
\beq
\label{geodn} \Delta N =N^i=0\ .
\eeq
Their motion in the comoving gauge \eqref{constphi} can then be found via the gauge transformation relating these gauges.  Specifically, in comoving gauge, the lapse and shift take the form\cite{Maldacena:2002vr}
\beq
N= 1 + {a\over {\dot a}} {\dot \zeta} + \calo(\zeta^2)\quad,\quad N^i = \partial_i \psi\ 
\eeq
with
\beq
\psi = -{\zeta \over a \adot} + {1\over \nabla^2} \left({\phidot_0^2 a^2 {\dot \zeta}\over 2\adot^2}\right)\ .
\eeq
So, the gauge transformation \eqref{lingauge} from \eqref{geodn} to comoving gauge is given by
\beq
\epsilon^0=\int dt {a\over \adot}{\dot \zeta}
\eeq
and
\beq\label{stoc}
{\dot \epsilon}^i = - {\phidot_0^2 a^2 \over 2\adot^2} {\partial_i \over \nabla^2}{\dot \zeta} + {\partial_i \zeta\over a\adot} + h^{ij} \int dt  {a\over \adot}\partial_j{\dot \zeta}\ 
\eeq
and in the comoving gauge the satellite trajectories are ${x}^i = -{\epsilon}^i+$\,const.

In \eqref{stoc}, the last two terms redshift away.  In slow roll, with $\phidot_0^2/H^2 \simeq V^{\prime 2}/V^2 = 2\epsilon$, the displacement is thus dominated by
\beq
{\dot x}^i=-{\dot \epsilon}^i \simeq \epsilon {\partial_i\over \nabla^2} {\dot \zeta} \ ,
\eeq
and for constant slow-roll parameter,
\beq
x^i = \epsilon {\partial_i\over \nabla^2} {\zeta}+ {\rm const.}
\eeq
We can then evaluate the leading contribution to the displacement:
\beq
\left< (\delta x^i)^2\right> = {\epsilon^2\over 2(2\pi)^2} \int{dk\over k} d(cos \theta) {k_i^2\over k^4} {H^2\over 2\epsilon} \propto \epsilon H^2 \int {dk\over k^3}\ .
\eeq
This is dominated in the infrared, leading to the dependence  
\beq
\left< (\delta x^i)^2\right> \sim \epsilon H^2L^2\ ,
\eeq
and thus a typical small variation in separation given by $\delta L \sim {\sqrt \epsilon} H L \ll L$.

\subsubsection{The $\langle\zeta\rangle$ tadpole contribution}

The tadpole of $\gamma$ must vanish,  since it has to be invariant under 3-dimensional Euclidian transformations, and there is no such invariant, symmetric, traceless two-index tensor. However, one might still worry about possible IR divergent contributions from the tadpole of $\zeta$, which in the slow-roll case could lead to effects competing with those in {\it e.g.} eq.~(\ref{W0}). But if the splitting between background and fluctuations is properly handled, these effects will be subleading. This is most easily seen by first considering spatially-flat gauge, \eqref{zerozeta}. In this gauge, the inflaton field can be written in terms of its backgound v.e.v. and the inflaton field fluctuations
\beq
\phi=\phi_0 + \varphi
\eeq
where by definition 
\beq
\left< \phi\right> =\phi_0~,\qquad \left<\varphi\right> =0~.
\eeq
Now we can use the gauge transformation from spatially-flat gauge into comoving gauge \eqref{constphi} on super-horizon scales to second order (see \cite{Maldacena:2002vr})
\beq
\zeta = \zeta_n + \frac{1}{2}(2\ep-\eta)\zeta_n^2
\eeq
with $\zeta_n$ defined by the linear relation
\beq
\zeta_n\equiv -\frac{H}{\dot\phi_0}\varphi~.
\eeq
From the tadpole condition above, we then have $\left<\zeta_n\right> =0$, and to second order we obtain
\beq
\left<\zeta(x)\right> = \frac{1}{2}(2\ep-\eta)\left<\zeta^2(x)\right>~.
\eeq
Although the variance $\left<\zeta^2(x)\right>$ contributes an IR divergence to the tadpole, it is clearly slow-roll suppressed compared to the other IR divergent contributions in eq.~(\ref{W0}).

\subsubsection{Variance of the line integral}\label{VLI}

Let us consider the contribution from tensor fluctuations to the variance of the line integral in eq.~(\ref{stav}) around the torus in the slow-roll, since this is the most interesting case. Since the tadpole of the tensor modes vanishes, the variance is
\beq
\left<\Delta\cals \Delta \cals\right> = \frac{1}{4}a^2(t)\lim_{yz\rightarrow y'z'} \left<\int_0^Ldx\gamma_{xx}(t,{\bf{x}})\int_0^Ldx'\gamma_{xx}(t,{\bf{x}}')\right>~.
\eeq
When applying the boundary conditions for the torus, the momentum is discretized and we are projecting on the zero mode in the $x$-direction, obtaining 
\beq\label{svar}
\left<\Delta\cals \Delta\cals\right> = \frac{1}{4}a^2(t)L^2 \left(\frac{1}{L^3}\sum_{k\in 2d}\left|\gamma_k(t)\right|^2\right)~.
\eeq 

In de Sitter, we have  $\left|\gamma_k(t)\right|^2 \approx H^2/k^3$ for super-horizon modes, and the contribution to the variance becomes of order $\left<\Delta\cals \Delta\cals\right> \approx a^2L^2H^ 2$, which is small and not IR enhanced. On the other hand, in slow-roll, there is a small IR enhancement, due to the non-flat spectrum $\left|\gamma_k(t)\right|^2 \approx H_k^2/k^3 \approx (H^2/k^3)(k/aH)^{n_t}$, where $n_t=-2\epsilon$ and $\epsilon >0$. Here, we  write the expansion rate as a function of momentum at horizon crossing instead of time by integrating eq.(5.18) of \cite{Giddings:2010nc},
\beq
H_k = \sqrt{\frac{\lambda}{3}}\left[\phi_e^2-8\ln\left(\frac{k}{a_e H_e}\right)\right]~,
\eeq
where we assumed for simplicity a standard potential $V(\phi) =\lambda\phi^4$; ``$e$" denotes values at the end of inflation.  The sum \eqref{svar} is dominated by smallest $k$, and is of size
\beq
\frac{1}{L^3}\sum_{k\in 2d}\left|\gamma_k(t)\right|^2 \simeq c \frac{H_e^2}{\phi_e^4}\log^2\left({La_e H_e}\right) 
\eeq
where $c$ is an $\calo(1)$ constant.
In order to evaluate the size of the effect, we assume that slow-roll inflation started in our local pocket near the self-reproduction scale, such that  \cite{Giddings:2010nc}
\beq
\frac{H_e^2}{\phi_e^4}N^2 =\frac{1}{3}\lambda N^2 =\frac{\pi^2}{8N_*}\mathcal{P}^{1/3}_{\zeta}=\frac{\pi^2}{8N},
\eeq
where $N$ is the total number of efolds, $N_*\approx 60$ is the number of e-folds left of inflation when the observable modes exits the horizon, and $\mathcal{P}_{\zeta}=(2.43\pm 0.11)\times 10^{-9}$ is the observed amplitude of the primordial curvature perturbation spectrum. With these values one obtains $(H_e^2/\phi_e^4)N^2\sim10^{-5}$. Thus, the variance is of order $a^2L^2\times 10^{-5}$, which is still small.

\section{Volume fluctuations}

Another measure of spatial geometry in inflationary cosmologies is the volume; here we briefly comment on this as a diagnostic of fluctuations.  

First, in the flat gauge \eqref{zerozeta}, the tracelessness of $\gamma_{ij}$ implies that the local volume element $\sqrt{h} d^3x$ is just given by that in the unperturbed solution \eqref{classsoln}.  Indeed, vanishing fluctuation in the local volume element can be thought of as a ``physical" way of describing the origin of the choice of slicing \eqref{zerozeta}.

In contrast, when one chooses the time slicing specified by the comoving gauge \eqref{constphi}, the fluctuations in the scalar field $\phi$ specifying the slicing lead to significant variations in the local spatial volume element.  This effect has  been explored in some detail in \cite{Creminelli:2008es,Dubovsky:2008rf}, which, within the context of eternal inflation, specifically studied the probability for the volume on the reheating time slice, $\phi=\phi_{rh}$, to become infinite.
Specifically, the fluctuation contribution to the volume element at quadratic order is given by
\beq
\left\langle e^{3\zeta}\right\rangle = e^{{9\over 2} \left< \zeta^2\right>}\ ,
\eeq
and so
the average volume of the reheating surface can be approximated by  
\beq
\left<V(t_{rh})\right> = e^{3\left(N+\frac{3}{2}\left<\zeta^2(x)\right>\right)}\left<V(t_{in})\right>\ ,
\eeq
where $t_{rh}$ is the time of reheating, $t_{in}$ is the initial time when inflation commenced, and $N$ is the number of efolds of the unperturbed solution \eqref{classsoln}.

We can estimate when the effect on the volume at  reheating becomes large; this happens when $\left<\zeta^2(x)\right>$ becomes of order $2N/3$. In a simple model of chaotic inflation with a monomial potential $V(\phi)= \lambda\phi^\alpha$,  the total number of e-folds scales with the initial inflaton field value, $\phi_i$, as $N=\phi_i^2/(2\alpha)$, while the variance scales as $\left<\zeta^2(x)\right> = \lambda\phi_i^{\alpha+4}/(12\pi^2\alpha^3(\alpha+4))$\cite{Giddings:2010nc}. This implies that the condition  
$\left<\zeta^2(x)\right> \gtrsim 2N/3$ is satisfied when
\beq
\phi_i\gtrsim (4\pi^2\alpha^2(\alpha+4)/\lambda)^{\frac{1}{\alpha+2}}~,
\eeq
which is comparable to the inflaton field value at the end of the self-reproduction scale
\beq
\phi_{sr}\sim (12\pi^2\alpha^2/\lambda)^{\frac{1}{\alpha+2}}~,
\eeq
but smaller than the field value when the potential energy density becomes of order $\mathcal{O}(M_p^4)$,
\beq
\phi_p \sim \lambda^{-1/\alpha}~.
\eeq
Thus, large fluctuations in the volume are closely associated with self-reproduction, between the Planck scale and self-reproduction scale, $\phi_p>\phi>\phi_{sr}$, as described in \cite{Creminelli:2008es,Dubovsky:2008rf}.  In this regime, our perturbative approach breaks down \cite{Creminelli:2008es,Dubovsky:2008rf,Giddings:2010nc,Giddings:2011zd,Burgess:2010dd} and there is no clear approach to computing the full ``wavefunction of the universe."  However, below this, in the non-eternal regime, the effect on the volume is small compared to the effect described in section \ref{PiPL}.

\section{Extensions and Conclusions}

In describing the quantum state of the geometry, one needs to formulate gauge-invariant q-observables; quantities such as $\langle g_{\mu\nu}(x)\rangle$ or even $\langle{\cal R}(x)\rangle$ are not gauge invariant.  This paper has described one way of characterizing inflationary geometry, via nonlocal q-observables involving lengths of geodesics in the fluctuating spacetime.  In the case where space is a torus, and in a time slicing with zero local deviation from classical expansion, we have found that fluctuations in the length around a cycle can become significant, and indeed such a curve can become shorter than an unperturbed curve, fixed in the original comoving coordinates, by a fractional amount that grows with the variance in the tensor fluctuations.  This provides evidence that accumulation of these fluctuations is indeed an effect with physical consequences.  Moreover, such contributions become large by a time scale $t\sim 1/H^3$ (or $t\sim R_{dS} S_{dS}$ in terms of the de Sitter radius and entropy, in general dimensions).  This is consistent with the suggestion that a perturbative description of dS fails at longer times, and with instability of dS space as a classical geometry\cite{Giddings:2007ie,Giddings:2010nc,Giddings:2011zd}.

 We have thus found that one of the simplest formulations of gauge-invariant quantities receives large corrections due to quantum fluctuations.  
We might seek other characterizations of geometries, besides such geodesics, for better understanding of these and other effects.  The challenge of respecting diffeomorphism invariance appears significant.  For example, as noted, one may choose a path and examine the holonomy around that path, but the choice of path will not be diffeomorphism invariant.  While workable measures of differences between geometries are scarce, we mention two possibly worth further exploration.  The first is the Gromov-Hausdorf distance between two geometries.  This is based on finding an embedding of these into a common geometry, and calculating a minimum distance over all such embeddings.  A second is the DeWitt distance.  If we have a fixed time slicing, so consider a difference $\delta g= g_1 -g_2$ between spatial metrics,\footnote{One can also work directly in terms of the spacetime geometry.} this can be written
\beq\label{dwd}
|\delta g|^2 = \int d^3x \sqrt{g} g^{ij} g^{kl} (\delta g_{ik} \delta g_{jl} + K \delta g_{ij} \delta g_{kl})\ .
\eeq
Here $K$ is a constant; it may be chosen so that in a harmonic gauge, such as $\nabla^j(\delta g_{ij} -{1\over 2}g_{ij}\delta g)=0$, the linear gauge transformation \eqref{gendiff} has zero effect.  DeWitt distance also appears to grow with the variance.

If the inflationary spacetime is adorned with additional structure, say either quantum or classical matter, additional statements can be made.  Specifically, matter allows one to formulate other relational observables, as {\it e.g.} discussed in \cite{Giddings:2005id}.  One may consider, as in the example of this paper, an initial distribution of (approximately) classical objects, and describe evolving features of the geometry with respect to these.  

In the case where we try to describe the geometry in terms of lengths between such ``satellites," we find behavior that may have been unexpected.  For starters,
as the path endpoints reach separations large as compared to the Hubble scale $1/H$, there is no spacetime geodesic, even in the unperturbed geometry, along which one can define the length.  An alternative prescription is to define the length in terms of geodesics constrained to lie within spatial slices.  Different choices of such spatial slices include those determined by either of the conditions  \eqref{constphi} or \eqref{zerozeta}.  However, when one computes the lengths of the geodesics in the perturbed metric, one finds that the short-wavelength metric fluctuations lead to significant shortening of the geodesics, with an $\calo(1)$ contribution for geodesics of length $\cals\sim 1/H^3$ (and, in the case of \eqref{constphi}, shorter by a slow-roll factor $\epsilon$).  Thus, one appears to lose control of a perturbative description of the geometry given in terms of such quantities.\footnote{This might be avoided at a coarse-grained level by introducing additional structure giving a long-distance averaging prescription, as noted above.}

One may also try to formulate diffeomorphism-invariant correlators of local operators in terms of proper-distance separation, as in \eqref{corrFL} \cite{Hamber,Ambjorn:1996wc,Giddings:2005id,Urakawa:2010it,Urakawa:2010kr,Urakawa:2011fg,Byrnes:2010yc,Gerstenlauer:2011ti}.  Here the above surprises also play a roll, forbidding such a construction with spacetime geodesics longer than $\sim 1/H$, and producing large fluctuations in geodesics constrained to spatial slices, when the operators are separated by distances $\lesssim \calo(1/H^3)$ (or $R_{dS} S_{dS}$ in general dimension).  At longer length scales, characterization of correlators in terms of spatial geometry thus appears to encounter subtleties. While there are such nonlocal, diffeomorphism-invariant constructions that are sensitive to growth of fluctuations, there also do  appear to be alternative sensible approximately local observables, that in particular are infrared safe, for observations restricted to small regions; specifically, those can be defined with an effective resolution parameter that is the size of the horizon of a given observer, and apply to present-day observations\cite{Giddings:2011zd}. (See also \cite{HHH,HMM}.)

In short, when one attempts to describe the quantum state, and some of its observables, at sufficiently long scales, one finds a breakdown in perturbative calculations.  There is no known approach to accurately calculate this state, and this apparently requires nonperturbative gravitational dynamics.\footnote{A similar story appears present in black hole geometries, at times $t\sim RS$.  This supports the resolution of the information paradox proposed in \cite{Giddings:2007ie}.  This leaves an information problem, which requires understanding the long-distance modifications to the state resulting from the nonperturbative gravitational dynamics.}  One can nonetheless apparently calculate certain physical quantities in small enough regions, over small enough time intervals.

\vskip.1in
\noindent{\bf Acknowledgements} We wish to thank J. Hartle, A. Hebecker, D. Marolf, D. Morrison, L. Senatore, and M. Zaldarriaga  for discussions, and D. Marolf for comments on a draft of this paper. The work of SBG was supported in
part by the U.S. Dept. of Energy under Contract
DE-FG02-91ER40618, and by grant FQXi-RFP3-1008 from the Foundational Questions Institute (FQXi)/Silicon Valley Community Foundation.

\appendix

\section{Gauge fixing  to higher order}

For our discussion of second-order effects in the $H/M_p$ expansion, it is important to check gauge fixing to this order.  Let us parameterize the metric as
\beq
 ds^2= -N^2 dt^2 + h_{ij}(dx^i + N^idt)(dx^j + N^jdt)\quad ;\quad h_{ij} = a^2(t)(\delta_{ij} +\hat h_{ij})~.
\eeq
Under combined time and space diffeomorphisms,
\beq
t \to t'=t-\epsilon^0\quad ,\quad x^ i \to x^{i\prime}= x^i - \ep^i
\eeq
the metric transforms to second order as
\beq \label{gaugetr}
\hat h'_{ij} = \hat h_{ij}+\p_j\ep_i+\p_i\ep_j +2\ep^0H\delta_{ij} +\delta_2{\hat h}_{ij}\ ,
\eeq
where we expand $N-1$, $N^i$, ${\hat h}_{ij}$, and $\epsilon$ order-by-order in $H/M_p$, and $\delta_2{\hat h}_{ij}$ is a term quadratic in the first-order fields and gauge parameters.

For general $\hat h_{ij}$ (and lapse and shift), we want to find a gauge transformation so that 
\beq
\delta_{ij} +\hat h'_{ij} = [e^{\gamma'}]_{ij}
\eeq
with $\p_i\gamma'_{ij}=\gamma'_{ii}=0$.

From the  trace  of \eqref{gaugetr}, we get to first order
\beq\label{firstord}
\ep_0^{(1)} =\frac{1}{6H}(\hat h_{ii}^{(1)}+2\p_i\ep_i^{(1)})\ ,
\eeq
where $H=\dot a/a$,
and to second order 
\beq
\ep_0^{(2)}=\frac{1}{6H}\Biggl[
\hat h_{ii}^{(2)} +2\p_i\ep_i^{(2)} +\delta_2{\hat h}_{ii}
-\frac{1}{2}\gamma_{ik}^{(1)\prime}\gamma_{ki}^{(1)\prime}\Biggr]
\eeq 

Now taking the divergence of eq.~(\ref{gaugetr}) gives to first order, using \eqref{firstord},
\beq
 (\p^2\delta_{ij}+\frac{1}{3}\p_i\p_j)\epsilon_i^{(1)}=-(\p_i\hat h_{ij}^{(1)}-\frac{1}{3}\p_j\hat h_{ii}^{(1)})
\eeq
({\it c.f} \eqref{gTtrans}).  This determines $\gamma^{(1)\prime}_{ij}$ to first order, via \eqref{gaugexm} and \eqref{transfix}.

At second order, we find
\beq
(\p^2\delta_{ij}+\frac{1}{3}\p_i\p_j)\epsilon_i^{(2)}= {\cal F}_j \ ,
\eeq
where ${\cal F}_j$ is given in terms of the second order $\hat h$ and the first order $\hat h$ and $\epsilon$'s.  This can be inverted, as in \eqref{transfix}, to determine $\epsilon_j$ at second order.  Note that this procedure may be iteratively continued to higher order.

\end{document}